\documentclass[a4paper,11pt]{article}

\usepackage{jcappub} % for details on the use of the package, please
\usepackage{booktabs}
\usepackage{centernot}
\usepackage{graphicx}% Include figure files
\usepackage{dcolumn}% Align table columns on decimal point
\usepackage[dvipsnames]{xcolor}
\usepackage{bm}% bold math
\usepackage{hyperref}% add hypertext capabilities
\usepackage[mathlines]{lineno}% Enable numbering of text and display math
\usepackage{xcolor}
\usepackage{amsmath}
\usepackage[capitalize]{cleveref}
\usepackage[normalem]{ulem}
\usepackage{csquotes}
\usepackage{multirow}
\usepackage{xspace}
\usepackage{enumitem}
\usepackage{chngpage}
\usepackage{threeparttable}
\usepackage{tablefootnote}

\newcommand{\prym}{\texttt{Prymordial}\xspace}
\newcommand{\parth}{\texttt{Parthenope}\xspace}

\newcommand{\relpow}[3]{\ensuremath{\left(\frac{#1}{#2}\right)^{#3}}}

\abstract{We summarize and explain the current status of time variations of the electron mass in cosmology, showing that such variations allow for significant easing of the Hubble tension, from the current $\sim5\sigma$ significance, down to between $3.4\sigma$ and $1.0\sigma$ significance, depending on the precise model and data. Electron mass variations are preferred by Cosmic Microwave Background (CMB) data in combination with the latest results on baryonic acoustic oscillations (BAO) and type Ia supernovae at a level of significance between $2\sigma$ and $3.6\sigma$ depending on the model and the data. This preference for a model involving an electron mass variation is neither tightly constrained from light element abundances generated during big bang nucleosynthesis nor from post-recombination observations using quasars and atomic clocks, though future data is expected to give strong evidence in favor of or against this model.}

\begin{document}

\title{The mass effect -- Variations of the electron mass and their impact on cosmology}

\author[1]{Nils Sch\"oneberg,}
\affiliation[1]{Institut de Ciències del Cosmos (ICCUB), Facultat de F\'isica, Universitat de Barcelona (IEEC-UB), Mart\'i i Franqués, 1, E08028 Barcelona, Spain}
\author[2]{L\'eo Vacher}
%\email{lvacher@sissa.it}
\affiliation[2]{International School for Advanced Studies (SISSA), Via Bonomea 265, 34136, Trieste, Italy}

\date{\today}
\emailAdd{nils.science@gmail.com}
\emailAdd{lvacher@sissa.it}

\maketitle

\setlength{\parskip}{0.5\baselineskip plus 0.1\baselineskip}
\setlength{\parindent}{0pt}

\section{Introduction}

\enlargethispage*{2\baselineskip}
Our current understanding of cosmology relies heavily on our most fundamental theories of nature: quantum field theory (QFT) and general relativity (GR), describing particle physics and the gravitational interaction respectively. The outstanding accuracy at which the cosmological $\Lambda$CDM model fits our observational data thus provides a powerful validation that our most fundamental theories of physics -- designed to describe local phenomena on earth and nearby astrophysical objects -- can be extended as far as the very birth of our Universe.

However, the recent accumulation of theoretical and observational puzzles suggests that QFT and GR and in particular the $\Lambda$CDM cosmological standard model are not the end of the story we are trying to tell. Keeping only some of the most stringent examples, our model of particle physics will undoubtedly have to be extended in order to include the observed neutrino masses as well as very likely the dark matter, while GR will have to be extended to a quantum theory in order to describe extreme phenomena as black holes or the primordial singularity. 

The $\Lambda$CDM standard model does not only inherit these theoretical challenges (such as the nature of the cold dark matter or the cosmological constant), but is also plagued by increasing observational tensions in the cosmological parameters, such as the $H_0$ tension that has recently risen beyond $5\sigma$ significance \cite{Verde2023, Planck2018VI, Riess2022}. These shortcomings directly point towards the possible necessity to add to or modify the fundamental building blocks of our model \cite{DiValentino:2021izs}.

Within this big picture, fundamental constants arguably play a crucial and unique role, as keystones of our theoretical constructions. They indeed set the relative magnitudes of all fundamental processes. Moreover, as they can not be computed but only measured, the constants clearly draw the borders of the predictive power of our models. Indeed, representing the limit of our theoretical understanding, they provide natural starting places to begin the search for extensions of our prevalent models. As such, looking for their possible variations in space and time gives a powerful and model independent probe of new physics. Furthermore, most of the promising theories beyond our standard models, as string theories \cite{Damour2003} and grand unified theories (GUTs) \cite{Dine2003,Dent2004}, predict a space-time variation of the fundamental constants. The measurement of their values at different space-time points are thus a rare and privileged observable to constrain these theories. 

We show in \cref{tab:fundamental} the fundamental constants that are in principle required to describe GR and the standard model of particle physics (without neutrino masses), noting that out of the 19 constants, 2 are Higgs couplings and 9 of them are Yukawa couplings. This means that more than half of our constants are dedicated to giving particles their masses. This can be taken as an indication that our understanding of the generation of particle masses has yet to be furthered. 

Due to the intricate way in which they appear within the standard model, it can be very challenging to implement variations of the masses of particles in a consistent way from first principles. It is, for example, much easier to construct  models including space-time variations of gauge couplings -- such as the fine-structure constant $\alpha$ -- by coupling a field directly to the gauge Lagrangian (see e.g. \cite{bekensteinOriginal}). However, recent claims from cosmological observations seem to indicate that a variation of the electron mass in the early Universe could significantly alleviate the $H_0$ tension, providing one of the most compelling models to date in that direction~\cite{H0olympics,Hart:2021kad,Chluba2023,Lee2023,Hart:2019dxi}. Such models of an electron mass variation are further aided by the inclusion of curvature \cite{Sekiguchi2021}, a fact we explain in \cref{ssec:me_cmb}. As such, the variation of the electron mass is of high importance to cosmology currently and deserves to be re-evaluated. In this work, we intend to present an updated review of the status of the electron mass in this discussion in regard of the latest cosmological and local datasets.

The paper is organized as follows: \cref{sec:shift_me} details a model where the value of the electron mass deviates from its current laboratory value during and before recombination. We examine the phenomenological effects of such a model and derive the latest observational constraints on it. This analysis shows that current data suggest a preference for a variation of the electron mass on cosmological timescales. We detail the impact of the electron mass variation on various datasets in three independent subsections. We first discuss the Cosmic Microwave Background (CMB) data supplemented by Baryon Acoustic Oscillation (BAO) and type Ia supernovae data in \cref{ssec:me_cmb}. This discussion is followed by constraints based on Big Bang Nucleosynthesis (BBN) in \cref{ssec:me_bbn} and the various local laboratory and astrophysical measurements in \cref{ssec:me_local}. For each probe considered, we first discuss the theoretical details of how electron mass variations impact the given probe(s) and then discuss the status of current constraints using that probe.

In \cref{sec:physical} we dive into the topic of how such variations can in principle be derived from a more fundamental model. We further investigate how such physically realized models might be constrained from local and astrophysical data. Finally, we present our conclusions in \cref{sec:conclusion}.

\enlargethispage*{1\baselineskip}
Before diving into the specifics of electron mass variations, let us make a few general notes. In this work we are not concerned with spatial variations of the electron mass. Such spatial variations are expected to give similar effects to the ones of the fine-structure constant discussed in \cite{Smith:2018rnu}, but we leave a precise investigation for future work. Additionally, any space-time variation (including only temporal) of any fundamental constant would lead to a violation of the Einstein equivalence principle (EEP) \cite{Will1993,Damour2012}, one of the cornerstones of the general theory of relativity (we comment on this point in \cref{app:EEP}). 

In such a case, one would be forced to introduce new fundamental forces and/or describe gravity with a non-metric theory. As such, fundamental constants can provide a powerful bridge between particle physics and gravity. Finally, when looking at mass variations, we always consider an evolution of the mass such that $m(z)/m_0 \neq 1$ at $z>0$ where $m_0 = m(z=0)$ is the mass measurable in laboratory experiments today. We discuss the theoretical background of the variations of such dimensionful quantities in \cref{app:mass_phys}.

\section{Constant shift of the electron mass on cosmological scales}\label{sec:shift_me}

The simplest model for the time-variation of a particle's mass is given by a constant offset in the early universe followed by an instantaneous transition during a time at which such a transition is difficult or impossible to observe. This simplest model allows us to study the first order effect of such mass variations without being too concerned about the actual theoretical foundations of how such mass shifts would be accomplished. As such, by design, these models are phenomenological in nature but still give valuable insights as they allow tracking the most direct impact of such a quantity on all of the relevant cosmological observables. Therefore, the model we study can explicitly be described as an extension of the usual $\Lambda$CDM model (involving six standard parameters, such as $\{\Omega_m h^2, \Omega_b h^2, H_0,  \tau_\mathrm{reio}, A_s, n_s\}$) through an additional parameter $\Delta m_e/m_e = [m_{e, \mathrm{early}} - m_{e,0}]/m_{e,0}$, where $m_{e,0}$ is the electron mass measured in current laboratory experiments and $m_{e, \mathrm{early}}$ is the electron mass in the early universe, see also \cite{H0olympics,Hart:2021kad,Chluba2023,Lee2023,Hart:2019dxi}. Later on we also study models with additional curvature specified through the curvature density fraction $\Omega_k$ as well as models using additionally the Chevallier-Polarski-Linder (CPL) parameterization of dark energy ($w_0/w_a$), first introduced in \cite{doi:10.1142/S0218271801000822,2003PhRvL..90i1301L}.

In the standard $\Lambda$CDM model the only massive particles with a large impact on cosmological observables are the protons, neutrons, electrons, and the cold dark matter. We focus here on the electron mass, which has a significant impact on multiple cosmological observables -- despite being several orders of magnitude smaller than the others -- due to the key role it plays in the electromagnetic interactions. During the cosmological evolution of our Universe, it is possible to identify three separate epochs at which the value of the electron mass can be relevant for observations, which are discussed in \cref{ssec:me_cmb,ssec:me_bbn,ssec:me_local}. Below, we detail why these three eras are especially impacted by a change of the electron mass. 

In the earliest phases of cosmic history ($T \gg m_e$), the ambient kinetic energy of the propagating particles is much higher than the electron mass, making the electron effectively massless for all reactions at this time. 

This soon changes during BBN, as the universe undergoes the decoupling of weak interactions (around $T \sim 1\mathrm{MeV}$, $z \sim 10^9$). This decoupling and the subsequent $\beta^-$ decay of the neutrons is tightly related to the relation between the neutron mass excess and the electron mass. The heavier the electron is, the more constrained is also the final phase-space of decays, and thus the more unlikely is the decay. This suppression of the decay rate for higher electron masses directly corresponds to a higher neutron abundance and thus leads to observably higher yields of Helium nuclei compared to hydrogen nuclei. These impacts of the electron mass on BBN through modifications of nuclear rates are detailed in \cref{ssec:me_bbn}.

After BBN, the electrons can essentially be seen as non-relativistic. However, they still participate in Compton scattering with photons (or Thompson scattering at these low energies). The threshold of where such scatterings are efficient or inefficient directly relates to the photo-ionization energies of the first atoms. The electron mass appears as a fundamental scale in most atomic transitions, dictating the binding energy and energy levels of all atoms, including (most importantly for cosmology) also hydrogen. %As such, it directly scales the temperature scale at which recombination can take place. 
The temperature scale at which recombination can take place is directly proportional to $\alpha^2 m_e$ (see \cref{ssec:me_cmb_theory} below) with a correction only mildly dependent on cosmological and physical parameters. The resulting impact on the recombination redshift explains why the variation of the electron mass has a major impact on observables which rely on how long acoustic oscillations could propagate before recombination, such as the CMB or the BAO. We discuss this impact in \cref{ssec:me_cmb}.

At even later times until redshift $z \gtrsim 10$ we find the cosmic dark ages during which we currently have no precise electromagnetic observations (though this might change in the distant future with 21cm mapping \cite{Burns:2021pkx,2011MNRAS.411..955M,2019BAAS...51g.241P}), giving us no strong handle on electron mass variations during this time.

Shifts of the electron mass at the latest times are typically disfavored through high-redshift observations of electromagnetic transitions such as in quasars, as well as through current-day high-precision observations from atomic clocks and the weak equivalence principle. We comment on all of these later time probes in \cref{ssec:me_local}. As such, it is typically assumed that the electron mass has returned to its current-day value already at a redshift before such observations, for example during the dark ages.

\subsection{Electron mass and the CMB}\label{ssec:me_cmb}

We now focus on the impact of the electron mass variation on the CMB and the constraints we can infer from it. First, in \cref{ssec:me_cmb_theory} we discuss the theoretical background of how electron mass variations affect the CMB. Then, in \cref{ssec:me_cmb_measure} we present the landscape of current constraints considering the CMB data alone and in combination with BAO data from the dark energy spectroscopic instrument (DESI) \cite{DESI:2024uvr,DESI:2024lzq} and type Ia supernovae data from PantheonPLUS \cite{Scolnic:2021amr}.

\subsubsection{Theoretical background}\label{ssec:me_cmb_theory}

The primary effect of a model including a variation of the electron mass on the CMB is its impact on recombination, such as most importantly through the hydrogen energy levels \mbox{($E_i \propto \alpha^2 m_e$)} or the Thompson scattering rate ($\sigma_T \propto \alpha^2/m_e^2$). See also \cite[Eq.~1]{Hart:2017ndk} for a summary of the dependencies on the electron mass for different rates involved in the recombination problem. As we will show below, the impact of an electron mass variation on the hydrogen energy levels (and in particular the binding energy) is tightly connected to the sound horizon and therefore also the Hubble tension, while the impact on the Thomson scattering rate is particularly important for the CMB damping envelope. In order to discuss these impacts, we start with a simplified model of recombination.

The impact of an electron mass variation on recombination can be estimated by assuming chemical equilibrium (the so-called Saha equilibrium). Using $\mu_x$ to denote the chemical potential of species $x$, this implies $\mu_{e^-} + \mu_{p^+} = \mu_H$\,. The non-relativistic abundance of each particle can be written as \cite{mukhanov2005physical}:
\begin{equation}
    n_x = g_x (m_x T/2\pi)^{3/2}\exp[-(m_x-\mu_x)/T]~,
\end{equation} with mass $m_x$ and multiplicity $g_x$ for each species. From the abundances, it is possible to construct the ratio $n_e n_p/n_H$ for which the dependence on the chemical potential cancels out according to the chemical equilibrium condition. Defining the free electron fraction as $x_e = n_e/n_H = n_p/n_H$ (assuming charge neutrality), the expression $n_e n_p/n_H$ can be written as the Saha equation \cite{coles2003cosmology,weinberg2008cosmology,dodelson2003modern}:
%%\begin{widetext}
\begin{align}\label{eq:saha}
    \frac{x_e^2}{1-x_e} = \relpow{\Omega_b h^2}{0.022}{-1} \overline{T}^{3/2} &\relpow{m_H}{1.00784\mathrm{amu}}{} \relpow{1-Y_\mathrm{He}}{1-0.245}{}\relpow{1+z}{1400}{-3/2} \nonumber\\
    &\cdot \exp\left[ 41.5 - 41.4 \relpow{1+z}{1400}{-1} \cdot \overline{T}^{-1} \relpow{m_e}{511\mathrm{keV}}{} (137\alpha)^2\right]~,
\end{align}
%%\end{widetext}
with $\overline{T} = T_\mathrm{cmb}/(2.7255\mathrm{K})\simeq 1$. Here we have made the dependence on all cosmological variables (including $m_e$) explicit.\footnote{We have assumed the numerical values for $G$, $\hbar$, and $c$. While $\hbar$ and $c$ relate directly to aspects of atomic physics crucial for recombination, $G$ appears implicitly in the densities, for example in $\Omega_b h^2 = 8\pi G\rho_b/(3 (100 \mathrm{km/s/Mpc})^2)$, thus contributing to the background evolution. The variation of these unit-defining constants is discussed in \cref{app:mass_phys}.} Looking at the term in the exponent we find that the redshift of recombination is $z_* \sim 1400$.\footnote{This number can be obtained by noticing that for $z_*$ the two terms in the exponential function are balanced. At higher redshift the first term dominates, giving $x_e^2/(1-x_e) \gg 1$, and therefore $x_e \to 1$. At lower redshift the second term dominates, giving $x_e^2/(1-x_e) \ll 1$, and therefore $x_e \ll 1$. We see that the quoted $z_*$ marks the redshift of transition between the two regimes.} \Cref{eq:saha} allows us to appreciate that the leading-order effect of an electron mass variation is an overall shift of the ionization curve to higher redshift (earlier recombination) for higher electron masses at recombination. Explicitly, we see that $z_* \propto m_e$\,. In reality the reactions quickly depart from the chemical Saha equilibrium, but the physical intuition described above still holds for more advanced numerical calculations of recombination \cite{Seager:1999bc,Seager:1999km,Ali-Haimoud:2010hou,2013ascl.soft04017C}. In particular, while recombination codes predict $z_* \sim 1080$ for the redshift of recombination, the scaling $z_* \propto m_e$ from the simple Saha model remains intact.

Correspondingly, a varying electron mass impacts the sound horizon $r_s$ defined as 
\begin{equation}
    r_s = \int_0^{\eta_*} c_s \mathrm{d}\eta = \int_{z_*}^\infty \frac{c_s \mathrm{d}z'}{H}~,
\end{equation}
with $\eta_*$ being the conformal time corresponding to $z_*$\,, $c_s$ the sound speed of acoustic waves, and $H$ the Hubble parameter. Increasing the electron mass increases the redshift of recombination and thus decreases the sound horizon. In flat $\Lambda$CDM, this sound horizon is observable under the acoustic angle
\begin{equation}
    \theta_s = \frac{r_s}{r_A} = (1+z_*)\frac{H_0 r_s}{\int_0^{z_*} \mathrm{d}z' ~1/E(z')}~,
\end{equation}
where $E(z) = H(z)/H_0$ is the relative expansion rate and $r_A$ is the angular diameter distance. Since this angle is tightly constrained by observations \cite{Planck2018VI}, it is important that any increase in the Hubble constant is compensated perfectly by a decrease in the sound horizon in order to remain in agreement with observations. As outlined above, a variation of the electron mass can accomplish such a decrease. This aspect of this model has regained attention recently in~\cite{Chiang:2018xpn,Sekiguchi2021,Baryakhtar2024}.

The angular scale of the sound horizon is not the only part of the CMB that is impacted by this model, though. To gain an intuition of the impact of a model with an electron mass variation on the CMB more generally, we can look at the CMB angular scales. The CMB primarily constrains three angular scales $\theta_{s}$\,, $\theta_{\rm eq}$\,, and $ \theta_D$ \cite{1995ApJ...444..489H, 1995PhRvD..51.2599H, 1996ApJ...471..542H}, each corresponding to a given physical distance scale imprinted at recombination: the acoustic scale $r_s$ (the sound horizon), the matter-radiation equality scale $r_\mathrm{eq}$\,, and the diffusion damping scale $r_D$ (or Silk scale). For fixed physical density abundances ($\Omega_m h^2$ and $\Omega_b h^2$) the aforementioned three distance scales are roughly independent of the value of the reduced Hubble constant $h$ and late-time physics (such as curvature or dynamical dark energy). The angular scales are derived from the distance scales as $\theta_X = r_X/r_A$ for $X \in \{s, \mathrm{eq}, D\}$. They have been used to construct the distance priors \cite{2009ApJS..180..330K} adopted in the Wilkinson Microwave Anisotropy Probe (WMAP) analysis and have been measured with great precision by recent surveys mapping the CMB anisotropies such as Planck \cite{Planck2018VI,Chen:2018dbv}. Therefore any new physics in the early Universe will have to leave these angular scales invariant to agree with observations.

There are two more properties of the CMB, which are important to be left unchanged as they are tightly constrained. This is the photon-to-baryon ratio (related to the even-to-odd peak height ratio) and the Hubble rate (related to the heights of the peaks relative to the Sachs-Wolfe plateau), both evaluated at recombination \cite{Sekiguchi2021}. For the varying electron mass model it is possible to keep the three angular scales and these two additional properties invariant at the price of changing only three additional parameters (see below). This property is a distinguishing feature of a model including an electron mass variation, setting it aside from other models changing the redshift of recombination, such as ones involving a varying fine-structure constant \cite{Kaplinghat:1998ry, Avelino:2000ea, Battye:2000ds, Avelino:2001nr,Martins:2003pe, Martins:2010gu, Hart:2017ndk} or primordial magnetic fields \cite{Thiele:2021okz,Galli:2021mxk,Jedamzik:2023rfd}. The authors of \cite{Sekiguchi2021} have investigated how this is possible, and we reproduce some of their analytical arguments below.

First, by changing the density parameters $\Omega_b h^2$ and $\Omega_m h^2$ in a precise way for a given variation of $m_e$\,, it is possible to restore the photon-to-baryon ratio and the Hubble rate at recombination to their fiducial values. Using $\Delta_x$ as the fractional variation\footnote{In practice, these are defined as $\Delta_x = \ln(x/x_\mathrm{fid})$ for some chosen fiducial cosmology \enquote{$\mathrm{fid}$}, which for us will be the $\Lambda$CDM cosmology with parameters set from \cite{Planck2018VI}, though the precise values are not very important for the presented argument.}  of a given parameter $x$, this requirement implies (see \cite{Sekiguchi2021})
\begin{equation} \label{eq:CMB_prop_parameter_scaling}
    \Delta_{\Omega_b h^2} = \Delta_{\Omega_m h^2} = \Delta_{m_e}~.
\end{equation}
Interestingly, one can show that this entails for both the sound horizon and the matter-radiation equality scale that
\begin{equation}
    \Delta_{r_s} = \Delta_{r_\mathrm{eq}} = -\Delta_{m_e}
\end{equation}
In addition, due to the precise proportionality of the Thomson scattering cross section as $\sigma_T \propto \alpha^2/m_e^2$\,, it turns out that also the diffusion damping scale at recombination changes as
\begin{equation}\label{eq:diffusion_scaling}
    \Delta_{r_D} = -\Delta_{m_e}
\end{equation}
See \cite{Sekiguchi2021} for the detailed calculation.\footnote{We note that in their calculation they neglect non-equilibrium effects as well as changes in the Helium fraction. However, we confirmed through numerical tests that the statements are true at the 0.3\% level for $r_D$ for up to 10\% variations in the electron mass, at the $0.01\%$ level for $r_s$\,, and up to numerical precision for~$r_\mathrm{eq}$\,.} As such, we find that all three distance scales $r_X$ change in the same way. Therefore, the \emph{angular} scales $\theta_X$ can be left invariant by enforcing that the angular diameter distance $r_A$ should also change this way, such that in the ratio $\theta_X = r_X/r_A$ the changes cancel out. This requires
\begin{equation}\label{eq:angular_scaling}
    \Delta_{r_A} \stackrel{!}{=} \Delta_{r_s} = -\Delta_{m_e}~.
\end{equation}
The authors of \cite{Sekiguchi2021} have numerically determined that in flat $\Lambda$CDM this requires
\begin{equation}\label{eq:angular_scaling_parameters}
    \Delta_h = 3.23 \Delta_{m_e}.
\end{equation}
If all these parameter changes are done as prescribed by \cref{eq:angular_scaling_parameters,eq:CMB_prop_parameter_scaling}, all five CMB properties can be returned to their fiducial values while only changing the electron mass and three additional parameters. In this case, the CMB becomes almost entirely insensitive to such a shift of the electron mass \cite{Sekiguchi2021}. Additionally, \cref{eq:angular_scaling_parameters} implies that in this case the Hubble parameter can be increased, opening the possibility of resolving the Hubble tension without any negative observational consequences.

\color{black}

The given parameter variations of \cref{eq:CMB_prop_parameter_scaling,eq:angular_scaling_parameters} imply together that
\begin{equation}\label{eq:final_h_scaling}
    \Delta_{\Omega_m} = -5.46 \Delta_{m_e} = -1.69\Delta_{h}~.
\end{equation}
As such, solutions to the Hubble tension based on electron mass variations typically result in lower values of $\Omega_m$\,, which has been shown in \cite{Lee2023,Lynch2024} to be a potentially critical issue for these models. Indeed, very low values of $\Omega_m$ are in principle incompatible with supernovae and BAO data. However, there are two reasons for not immediately discarding this model as a solution to the Hubble tension.

First, it turns out that some new BAO observations from DESI \cite{DESI:2024mwx} prefer mildly lower values of $\Omega_m$ ($\Omega_m = 0.295 \pm 0.015$ in flat $\Lambda$CDM, and even lower with curvature). While supernovae typically prefer higher values (PantheonPLUS has $\Omega_m = 0.334\pm 0.018$, Union3 has $\Omega_m = 0.356 \pm 0.027$, and DESY5 has $\Omega_m = 0.352 \pm 0.017$), we show below that at least for PantheonPLUS supernovae the constraint is apparently not tight enough to compete with BAO data in order to exclude such variations. Investigations of the impact of the other supernovae datasets are left to future work.

Second, there are simple modifications of the baseline model that diffuse such late-time constraints. For example, solutions have been proposed in
\cite{H0olympics,Khalife2024}, in which the inclusion of additional curvature $\Omega_k$ weakens the constraining power of these additional probes on~$\Omega_m$\,. Below, we show that a model with dynamical dark energy (using the CPL model with \mbox{$w = w_0 + w_a~\frac{z}{1+z}$}) shows the same kind of behavior of weakening the constraining power of late-time probes such as BAO or type Ia supernovae on electron mass variations. However such dynamical dark energy models are more susceptible to be strongly excluded with tight supernovae data. A second effect of these additional late-time modifications is that the angular diameter distance can be scaled like \cref{eq:angular_scaling} without requiring the parameter $h$ to be used for this scaling (as was assumed in \cref{eq:angular_scaling_parameters}). Instead, the additional parameters $\Omega_k$ and $w_0/w_a$ can be used for that purpose as well, leaving even more freedom in the choice of~$h$.

\subsubsection{Current status of measurements}\label{ssec:me_cmb_measure}

\begin{figure}
    \centering
    \includegraphics[width=0.6\textwidth]{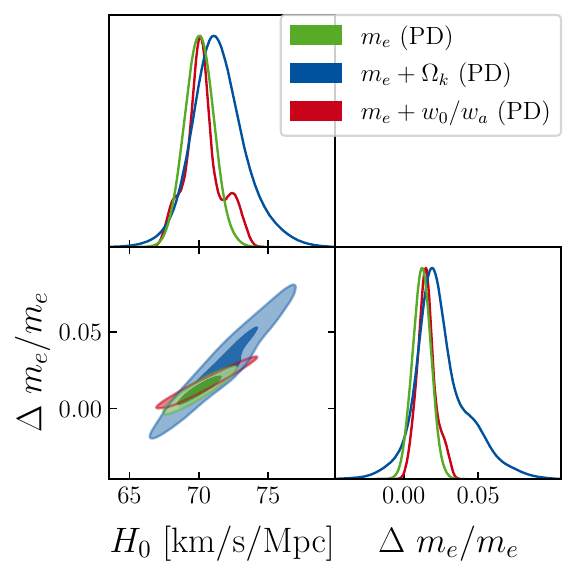}
    \includegraphics[width=0.6\textwidth]{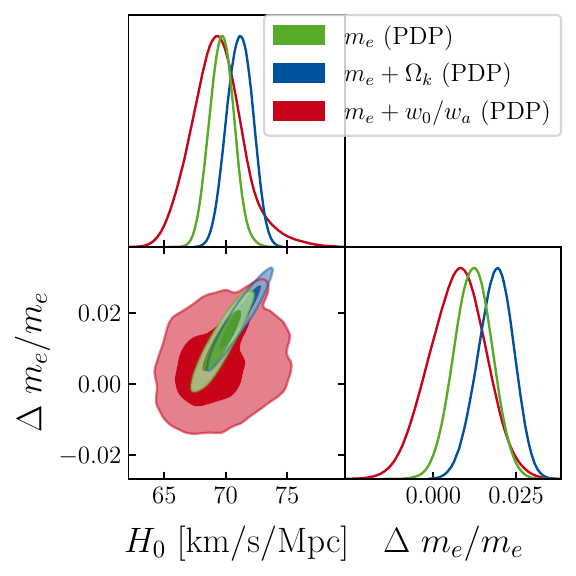}
    \caption{Constraints on variations of the electron mass ($m_e$), also with additional variation of curvature ($\Omega_k$) or CPL-dark energy ($w_0/w_a$). We show the 68\% and 95\% confidence contours an marginalized posteriors for P18+DESI data (PD) and also PanPLUS data (PDP), see \cref{tab:electron_mass_cmb} for references.}
    \label{fig:electron_mass_constraints}
\end{figure}

Naturally, the full extent of modifications to the CMB due to a variation of the electron mass can best be appreciated when following the calculations of a Einstein-Boltzmann solving code (such as \texttt{class} \cite{2011JCAP...07..034B}). We use the default implementation of varying constants in \texttt{class} for the results we show below, using the MCMC sampler implementation of \texttt{MontePython} \cite{Audren:2012wb,Brinckmann:2018} for running the chains and we produce plots using \texttt{liquidcosmo}.\footnote{The \texttt{liquidcosmo} code is based on \texttt{getdist} \cite{Lewis:2019xzd} and is available at \url{https://github.com/schoeneberg/liquidcosmo}.} We put flat priors on the seven model parameters introduced in \cref{sec:shift_me}, as well as on additional curvature or dynamical dark energy parameters.

As detailed in the previous subsection, a constant shift of the value of the electron mass provides a potential path towards alleviating the Hubble tension. As such, this scenario has been frequently investigated in the literature, using different datasets and different combinations of additional parameters. The current status of measurements of the electron mass from the CMB and BAO (without BBN) is summarized in \cref{tab:electron_mass_cmb} and our contributions are shown in \cref{fig:electron_mass_constraints}. The uncertainties are typically at the level of a few times $10^{-2}$ without BAO data, while they decrease to a little less than $10^{-2}$ with BAO data. However, this tighter constraint can be loosened by including additional curvature or additional dynamical variation of dark energy. We note that the latter is more sensitive to Pantheon supernovae data, as we show in table \cref{tab:electron_mass_cmb} and in \cref{fig:electron_mass_constraints}.

\begin{figure}
    \centering
    \includegraphics[width=0.6\textwidth]{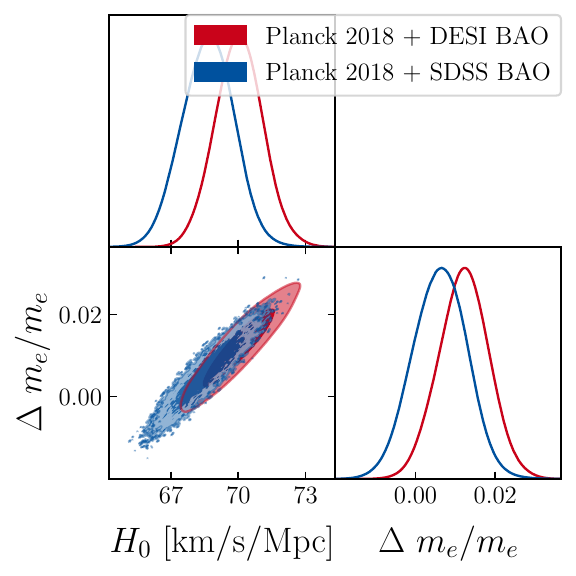}
    \caption{68\% and 95\% confidence contours an marginalized posteriors for variations of the electron mass, comparing the older SDSS BAO data to the newer DESI BAO data, see \cref{tab:electron_mass_cmb} for references.}
    \label{fig:electron_mass_oldbao}
\end{figure}

We also note that each case investigated in this work supports a positive variation of the electron mass anywhere between 1$\sigma$ and 3$\sigma$ significance, even without adding any prior on the Hubble parameter. We note that this preference appears to at least partially be driven by the DESI BAO data with its preference for a lower $\Omega_m$ value, which is compatible with a larger electron mass from \cref{eq:final_h_scaling}. We show evidence for this trend in \cref{fig:electron_mass_oldbao}, which presents an evident shift in the variation of the electron mass from $\Delta m_e/m_e = 0.0061 \pm 0.0068$ ($<1\sigma$ preference) for SDSS BAO to $\Delta m_e/m_e = 0.0121\pm0.0063$ ($\sim 2 \sigma$ preference) for DESI BAO. The highest significance is found for the $m_e+\Omega_k$ model with Planck, DESI, and PantheonPLUS data at a level of $\Delta m_e/m_e = 0.0188 \pm 0.0052$, which prefers a variation in the electron mass at the 3.6$\sigma$ level of significance. At the same time those models with a preference for an electron mass variation also predict a higher Hubble constant (even without a prior from \cite{Riess2021} that would drag the constraints in such a direction). When incorporating DESI data, the Hubble tension is reduced below $2 \sigma$ (see \cref{tab:electron_mass_cmb}). The lowest significance of the Hubble tension ($1.0 \sigma$) is reached when considering the $m_e+\Omega_k$ model with Planck, DESI, and PantheonPLUS data.

\begin{table*}[t]
    \centering
    \resizebox{\columnwidth}{!}{\begin{tabular}{l|c c c c c}
        Model & Data & $10^2 \Delta m_e/m_e$ & $H_0$ [km/s/Mpc] & GT & Reference \\ \hline
        $\Delta m_e/m_{e}$ & P15 & $-3.9^{+4.6}_{-7.2}$ & $60^{+8}_{-16}$ & 1.6 &\cite{Hart:2017ndk} \\
        $\Delta m_e/m_e$ & P15 + SDSS  & $0.39^{+0.74}_{-0.74}$ & $68.1\pm 1.3$& 3.0 & \cite{Hart:2017ndk} \\
        $\Delta m_e/m_e$ + $\Delta \alpha/\alpha$ & P15 + SDSS   & $0.56 \pm 0.80$ & $68.1\pm1.3$ & 3.0 & \cite{Hart:2017ndk}\\ \hline
        $\Delta m_e/m_e$ & P18  & $-11.2\pm 5.9$ & $46^{+9}_{-10}$ & 3.0 & \cite{Hart:2019dxi}, see also \cite{Baryakhtar2024}\\
        $\Delta m_e/m_e$ & P18 + SDSS  & $0.47 \pm 0.66$ & $68.46\pm 1.26$ & 2.8  & \cite{H0olympics}, see also \cite{Hart:2019dxi,Baryakhtar2024} \\
        $\Delta m_e/m_e$ + $\Omega_k$ & P18 + SDSS  & $1.5 \pm 1.8$ & $69.29 \pm 2.11$ & 1.6 & \cite{H0olympics}, see also \cite{Toda:2024ncp} \\ \hline
        $\Delta m_e/m_e$ & SPT+P18+SDSS+Pan & $0.3 \pm 0.6$ & $68.0\pm 1.1$ & 3.3 & \cite{Khalife2024} \\
        $\Delta m_e/m_e$ + $\Omega_k$ & SPT+P18+SDSS+Pan & $0.35 \pm 1.64$ & $68.2 \pm 1.6$&  2.5&\cite{Khalife2024} \\
        $\Delta m_e/m_e$ + $\Omega_k$ + $m_\nu$ & SPT+P18+SDSS+Pan & $3 \pm 3$ & $69.8^{+1.8}_{-2.9}$ & 1.6 & \cite{Khalife2024} \\ \hline
        $\Delta m_e/m_e$ & P18+DESI+SDSS+Pan & 
        $0.92 \pm 0.55$ & $69.44 \pm 0.84$ & 2.7 & \cite{Seto2024}\\
        $\Delta m_e/m_e + \Omega_k$ & P18+DESI+SDSS+Pan & $1.3 \pm 1.4$ & $69.7\pm1.4$ & 1.9 & \cite{Seto2024} \\ \hline
        $\Delta m_e / m_e$ & P18+Pan & $0.69^{+1.44}_{-1.36}$ & $68.75^{+3.01}_{-2.81}$ & $1.4$ & \cite{Baryakhtar2024}\\ 
        $\Delta m_e / m_e$ & P18+PanPLUS & $-1.00^{+1.09}_{-1.04}$ & $65.08^{+2.18}_{-2.03}$ & $3.4$ & \cite{Baryakhtar2024}\\ \hline
        $\Delta m_e/m_e$ & P18+DESI & $1.21 \pm 0.63$ & $70.03 \pm 1.06$ & 2.0 & this work \\
        $\Delta m_e/m_e + \Omega_k$ & P18+DESI & $2.52 \pm 1.85$ & $70.94 \pm 1.23$ & 1.3 & this work, see also \cite{Baryakhtar2024} \\
        $\Delta m_e/m_e + \Omega_k$ & P18+DESI+PanPLUS & $1.88 \pm 0.52$ & $71.61 \pm 1.00$ & 1.0 & this work \\
        $\Delta m_e/m_e + w_0/w_a$ & P18+DESI & $1.53 \pm 0.63$ & $70.30 \pm 1.34$  & 1.6& this work \\
        $\Delta m_e/m_e + w_0/w_a$ & P18+DESI+PanPLUS & $0.72 \pm 0.84$ & $69.38 \pm 2.17$ & 1.5 &this work
    \end{tabular}}
    \caption{Review of literature constraints on the simplest implementations of a constant cosmological shift of the electron mass as well as our own contributions of \cref{fig:electron_mass_constraints}. P15 refers to Planck 2015 CMB data (see \cite{Planck2015XIV}), while P18 refers to Planck 2018 CMB data (see \cite{Planck2018V}). SDSS refers to BAO during the BOSS/eBOSS era (before DESI) from the SDSS, such as DR12 LRG \cite{BOSS:2016wmc}, DR14 QSO/Ly$\alpha$ \cite{eBOSS:2019qwo,eBOSS:2019ytm}, and DR7 MGS \cite{Ross:2014qpa}, as well as small redshift BAO from 6dFGS \cite{2011MNRAS.416.3017B}. Pan refers to Pantheon uncalibrated supernovae\protect\footnotemark~from \cite{2018ApJ...859..101S} and PanPLUS to PantheonPLUS uncalibrated supernovae from \cite{Scolnic:2021amr}. Finally, DESI denotes data from \cite{DESI:2024uvr,DESI:2024lzq}. GT refers to the Gaussian tension with \cite{Riess2022} (as a rough relative indicator). If a massive scalar field is assumed to generate the electron mass variations, its cosmological impact is typically negligible, as discussed in \cite{Baryakhtar2024}.} \label{tab:electron_mass_cmb}
\end{table*}

\newpage 
\subsection{Electron mass and BBN} \label{ssec:me_bbn}

In this section we discuss whether the addition of constraints coming from light element abundances generated during BBN (preceding the emission of the CMB) could provide a major roadblock to the possible alleviation of the $H_0$ tension by electron mass variations.
First, in \cref{ssec:me_bbn_theory} we discuss how electron mass variations affect BBN, and in \cref{ssec:me_bbn_measure} we show the current status of constraints from BBN.

\footnotetext{\enquote*{Uncalibrated} in this context implies that one does not use any calibration of the supernovae absolute magnitude, such as for example the one from Cepheids \cite{Riess2022} or from the tip of the red giant branch \cite{2021ApJ...919...16F}.}
\enlargethispage*{1\baselineskip}
\subsubsection{Theoretical background} \label{ssec:me_bbn_theory}
It is well established that elements with higher atomic numbers have to be sequentially constructed from lighter elements \cite[p.~165]{weinberg2008cosmology}, as multi-particle fusions are disfavored by the extremely small baryon-to-photon ratio $\eta_b \simeq 6.1 \cdot 10^{-10}$. Correspondingly, the creation of the lightest heavy nucleus (Deuterium) is a Bottleneck for the creation of further heavier elements \cite{Iocco:2008va}. Similar to \cref{eq:saha} we can also create simplified prediction for the Deuterium generation in chemical equilibrium as
%\begin{widetext}
\begin{align}
    \frac{x_D}{x_p x_n} = & \relpow{\eta_b}{6.1\cdot 10^{-10}}{} \,  \, \overline{T}^{3/2} \, \mathfrak{m}^{3/2} \, \relpow{1+z}{2.74 \cdot 10^8}{3/2} \nonumber\\ &~\mbox{}\quad~\cdot \exp\left[-34.55
    + \overline{T}^{-1} \cdot 34.55 \relpow{1+z}{2.74\cdot 10^8}{-1} \relpow{B_D}{2.22452\mathrm{MeV}}{} \right]~,
\end{align}
%\end{widetext}
which predicts a redshift of $z \simeq 2.74 \cdot 10^8$, corresponding to a photon temperature of 64 keV for BBN, not far from the roughly $\sim 100$ keV commonly cited \cite{mukhanov2005physical,weinberg2008cosmology,Iocco:2008va,Fields2023,Cyburt:2015mya}. Here too we have kept all constants as above and have defined for convenience
\begin{equation}
    \mathfrak{m} = \relpow{m_D}{931.494\mathrm{MeV}}{} \cdot \relpow{m_p}{938.272\mathrm{MeV}}{-1} \cdot \relpow{m_n}{939.565\mathrm{MeV}}{-1}~.
\end{equation}
We observe that the time at which the Deuterium bottleneck is overcome is strongly dependent on the binding energy of Deuterium, which is mostly a quantity depending on strong nuclear forces, and therefore almost entirely independent of the electron mass. 

However, as mentioned in the introduction of \cref{sec:shift_me}, the crucial process to consider here is $\beta^-$ decay. Since the weak scatterings have frozen out at BBN (100 keV $\ll $ 1 MeV), the neutron number declines mostly only through $\beta^-$ decay. As such, there can only be as much Helium produced as there are available neutrons. Indeed, the efficiency of Helium fusion is so great that to first order it can be directly estimated from the abundance of neutrons, namely 
$Y_p = 4 n_\mathrm{He}/n_b \simeq 2 n_n / (n_n+n_p)$. The neutron-to-proton ratio is easily estimated using their initial abundance at the freezeout of weak interactions multiplied by the subsequent decay. Both the weak rate that determines their initial abundance and the decay rate are dependent on the electron mass. An increase of the electron mass results (for constant neutron-proton mass difference) overall in a decrease of the decay phase space and therefore the decay rate \cite{Burns:2023sgx,Seto2023}, thus more neutrons at the Deuterium bottleneck, leading therefore to an overall higher Helium abundance.

\subsubsection{Current status of measurements} \label{ssec:me_bbn_measure}
Overall, \cite{Seto2023} argues that the BBN constraint on the variation of the electron mass should be negligible compared to Planck CMB constraints in the standard model without curvature or dynamical dark energy, but they do not show constraints from BBN alone. Here we derive the constraints from BBN using the following approach:
\begin{enumerate}
    \item We modify the publicly available \prym code \cite{Burns:2023sgx} to accept the nuclear rates from the \parth code \cite{2022CoPhC.27108205G} -- In principle the computations could have been performed with any set of nuclear rates, but for the purpose of comparison with the rates used by default in \texttt{class}, we chose this set of rates (see also \cite{Schoneberg:2024ifp}).
    \item We modify the $m_e$ input parameter of the code, making sure to set the flag to recompute the weak rates for each computation. Since this significantly slows down the code, we chose to build an emulator (see below) to accelerate inference.
    \item We evaluate a random sampling in $\{\eta_{b}\,, \Delta m_e/m_e\}$ and record the predicted element abundances and the baryon fraction for each parameter point.
    \item With these predictions we build a linear interpolator in the two-dimensional parameter space that achieves an average emulation accuracy around $\sim 0.1\%$ for a range of test cases.
    \item We write a likelihood using that emulator in order to compare observed element abundances with emulated/predicted ones. We use that likelihood in the framework described in \cref{ssec:me_cmb_measure}.
\end{enumerate}

\begin{figure}
    \centering
    \includegraphics[width=0.7\textwidth]{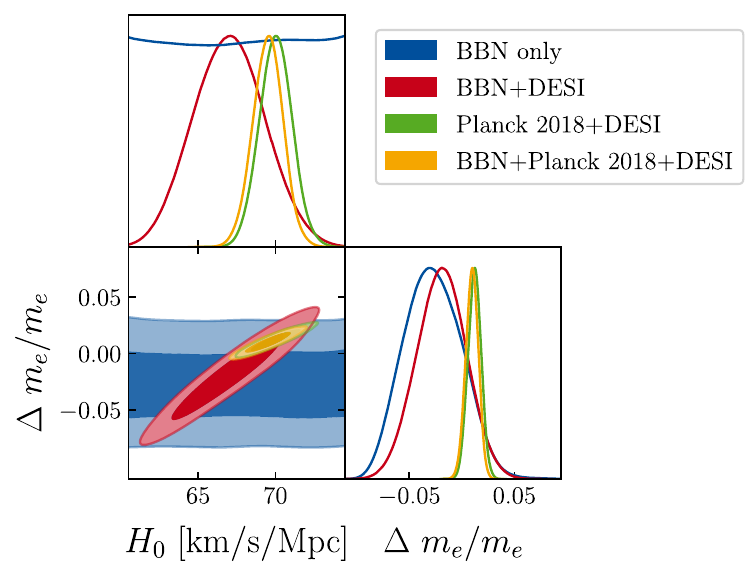}
    \caption{\label{fig:electron_mass_bbn} Same as \cref{fig:electron_mass_oldbao}, but involving data combinations with and without BBN.}
\end{figure}

The data we use are the particle data group (PDG) recommended abundances for $Y_\mathrm{He}$ and $D/H$ from \cite{ParticleDataGroup:2022pth} -- $Y_\mathrm{He} = 0.2450 \pm 0.0030$ and $D/H = (2.547 \pm 0.029) \cdot 10^{-5}$. We find the constraints as summarized in \cref{fig:electron_mass_bbn}. With only BBN data the constraint on $\Delta m_e/m_e$ is very wide, up to 5\% variation are allowed (whereas current 68\% CL constraints from the CMB are closer to around $\sim1\%$), in qualitative agreement with the arguments of \cite{Seto2023}. This loose constraint persists once BAO data are added, as these do not further measure the impact of an electron mass shift beyond a change in the sound horizon. 

As discussed in great detail in \cite{Schoneberg:2019wmt, Schoneberg:2022ggi, Addison:2013haa, BOSS:2014hhw, Addison:2017fdm,eBOSS:2019qwo,Cuceu:2019for} the combination of BAO and BBN leads to constraints on $H_0$\,, here at the level of $H_0 = 67.0 \pm 2.2$ km/s/Mpc. Since BAO measurements put tight constraints on $H_0 r_s$ and $\Omega_m$ (as well as other late-time parameters such as $\Omega_k$ or $w_0/w_a$), together with the determination of $\Omega_b h^2$ and $m_e$ from BBN the value of $r_s$ can be calibrated and a corresponding measurement of $H_0$ be obtained~\cite{Schoneberg:2022ggi}. The uncertainty of this determination of $H_0$ is then given by how strongly BBN data can constrain the impact of an electron mass variation on the sound horizon. The uncertainty is larger than in \cite{Schoneberg:2022ggi} due to the additional freedom of varying the electron mass and the corresponding shifts in the sound horizon, which we have discussed in \cref{ssec:me_cmb_theory}. In principle, as shown in \cite[Fig.~4]{Schoneberg:2021qvd} and \cite[Sec.~3.8]{Schoneberg:2022ggi}, varying such a parameter could completely degrade the constraining power, except if the additional parameter is also constrained by BBN measurements (as it is in this case).

Once CMB data are added back, the constraint becomes much tighter, though as before, non-zero values of the electron mass variation are preferred. The BBN does not add a lot of information beyond the Planck CMB constraints in this case, and only marginally shifts the constraint on $H_0$\,. 

\begin{figure}
    \centering
    \includegraphics[width=0.5\textwidth]{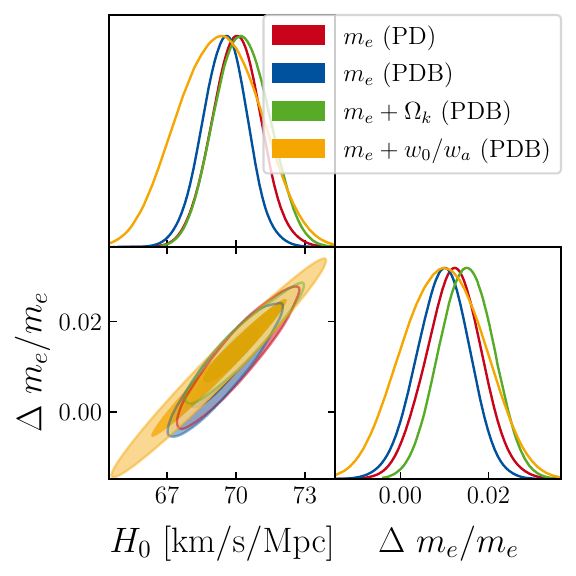}
    \caption{\label{fig:electron_mass_bbn_variations} Same as \cref{fig:electron_mass_bbn}, involving various models and data sets, with the same notation as in \cref{fig:electron_mass_constraints,tab:electron_mass_cmb} and PDB including P18 CMB data, DESI BAO data, and BBN abundances from~\cite{ParticleDataGroup:2022pth}.}
\end{figure}

We also show additional variations of the model with curvature or dynamical dark energy in \cref{fig:electron_mass_bbn_variations}. BBN can slightly help constraining these more exotic variations. For example for the case of curvature $\Omega_k$ it improves the constraint from $\Delta m_e/m_e=0.0252\pm0.0185$ without BBN to $\Delta m_e/m_e=0.0150\pm0.0055$ with BBN, which still shows an almost $3\sigma$ preference for an early electron mass variation. However, the broadest constraint of $H_0$ with this data combination including BBN is given by the case of $w_0/w_a$ CPL dark energy with $H_0 = 69.3 \pm 1.7$ km/s/Mpc ($1.9\sigma$ tension with \cite{Riess2022}).

Given the clear picture that neither BBN nor CMB can strongly constrain the electron mass variation (in such a way as to prevent it from reducing the Hubble tension), and that there even is a preference for electron mass shifts in the data, a natural question arises about other constraints on the variation of the electron mass.

\subsection{Electron mass and the post-recombination Universe\label{ssec:me_local}}

Having shown a preference for a shift in the electron mass as well as a reduction of the Hubble tension using CMB and BAO/type Ia supernovae data, and having demonstrated that such a preference cannot be ruled out by BBN abundance data, we now discuss if constraints in the post-recombination Universe are relevant for this model.
First, in \cref{ssec:me_local_theory} we discuss how electron mass variations affect post-recombination astrophysical and laboratory-based experiments, and in \cref{ssec:me_local_measure} we show the current status of constraints from these experiments.

\subsubsection{Theoretical background} \label{ssec:me_local_theory}

There are mainly two post-recombination types of measurements currently considered for constraining mass variations: astrophysical and laboratory-based. The first type of experiment measures the spectra of astrophysical objects at finite redshift. Typically, one would be interested in the absorption lines generated by cold and low density clouds in the line of sight towards bright and distant quasi stellar objects (QSO). Since the redshift of such objects is typically determined using spectral features, this measurement compares frequencies of different atomic lines. As such, the immediate dependence on the overall electron mass cancels out. For example, the Lyman-$\alpha$ transition compared to the Lyman-$\gamma$ transition always have a precise frequency ratio of $4/5$. 

Instead, one uses the fact that additional corrections to energy levels, such as from rotation, vibrational, and electronic degrees of freedom (mostly for molecules) depend on ratios that are (fractional) powers of $\mu \equiv m_p/m_e$ \cite{Jansen:2013paa,Ubachs2007,Uzan2011,Martins2017review}. For example, vibrational modes typically scale as $\mu^{-1/2}$ and rotational modes as $\mu^{-1}$ \cite{Jansen:2013paa,Karshenboim:2000rg}. As such, the comparison of different types of lines allow for constraints on the mass ratio $\mu$. The precise lines used are optimized such that their dependence on the mass ratio is enhanced. For example, if two different types of energy levels are very close to each other for the standard value of $\mu$, the relative change of their transition frequency will be substantially more sensitive to a given non-standard value of $\mu$. Explicitly, \cite{Ubachs2007} shows that if an electronic and the $n$-th vibrational energy levels are approximately equal, the mass-dependence of their transition is enhanced by the inverse of the small remaining transition frequency.

The second type of measurements are performed locally in laboratory environments using atomic clocks. It is possible to compare optical transitions with those of a hyperfine baseline. Given that the hyperfine transition depends on $\mu$ while the optical transition does not, a drift in their relative rate during a long-time laboratory experiment would signal a possible shift in $\mu$. Similar to the high-redshift measurements, different combinations of different kinds of transitions can provide different sensitivities to shifts in the mass ratio. However, many of the used transition frequencies also depend on the proton gyromagnetic ratio ($g_p$), which can either be assumed to be fixed to the standard model value or conservatively marginalized over.

Finally, the study of the universality of free fall can provide stringent bounds on possible mass variations, but the translation of the measurements into bounds on mass variations is not trivial. We discuss constraints from free-fall experiments in \cref{app:EEP}. 

\subsubsection{Current status of measurements} \label{ssec:me_local_measure}

The current status of late time measurements of possible shifts in the proton-electron mass-ratio $\mu$ using QSO can be found in \cite[Tab.~2]{Martins2017review}, and are typically of the order of around $10^{-6}$ to $10^{-5}$ for $z\in [0.685,4.224]$ hinting at a possible variation with a small ($\sim 1 \sigma$) level of significancy \cite[Eqs. (16) and (17)]{Martins2017review}. Since the highest observed redshift is around $4.2$, this means that any transition of the electron mass occurring in the dark ages would not be sensitively picked up by current experiments. 

Beyond these quasar constraints there are also tight laboratory constraints on the variation of the electron mass from optical atomic clock transitions, comparing an atomic transition ($\mu^{0}$) with a hyperfine Cs137 transition ($\mu^{-1}$), which is also used for the definition of the second, see the discussion in \cref{app:mass_phys}. 
We explicitly use the clocks from \cite[Tab.~3]{Martins2017review} 
(i.e.~\cite{doi:10.1126/science.1154622,2013PhRvL.111f0801L,Shelkovnikov:2008rv,2004LNP...648..209F,2015CRPhy..16..461A,Fortier:2007jf,2014PhRvA..89b3820T,2014PhRvL.113u0802H,Guena:2017lbc}) as well as additionally \cite{Lange:2020cul,Filzinger:2023zrs}.

\begin{figure}
    \centering
    \includegraphics[width=0.5\textwidth]{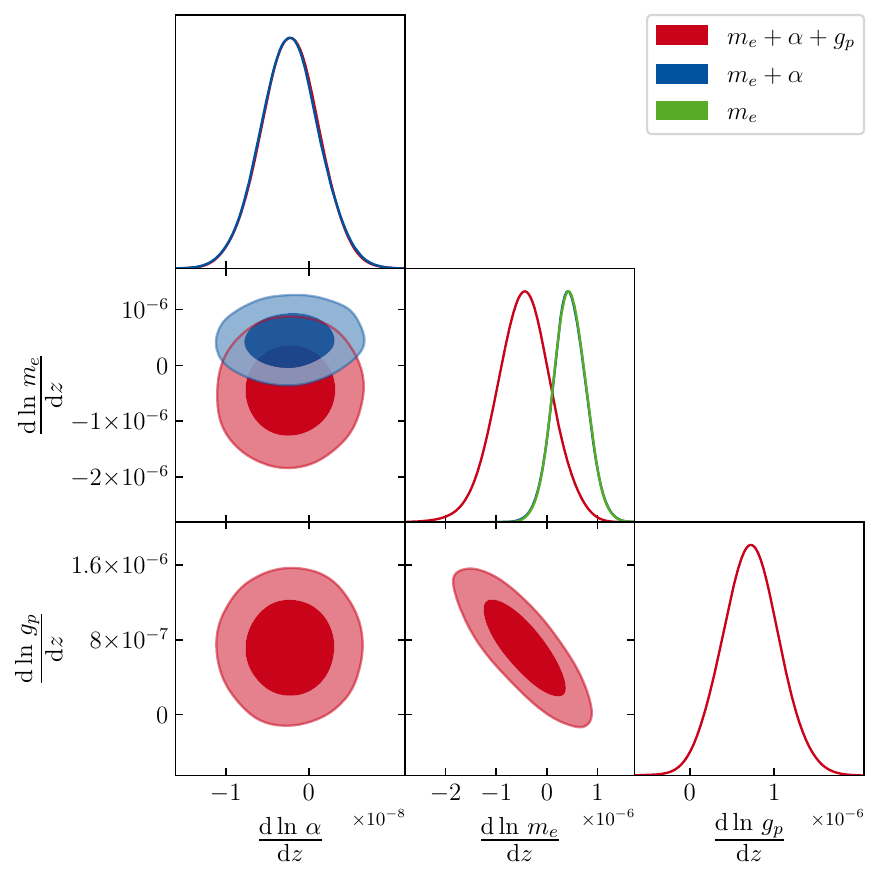}
    \caption{68\% and 95\% confidence contours an marginalized posteriors for atomic clock constraints (data described in main text) on the variation of rates for a few models with different variations of underlying parameters. Note that the scale of the axis for ${\rm d}\ln\alpha/{\rm d}z$ is two orders of magnitude smaller than for ${\rm d}\ln m_e/{\rm d} z$.}
    \label{fig:me_atomic}
\end{figure}

The resulting posteriors distributions of the time drifts of $\{\alpha, \mu, g_p\}$ are shown in \cref{fig:me_atomic}, leading to constraints of a few times $10^{-7}$ in $\mathrm{d}\ln m_e/\mathrm{d}z$ after marginalization. While very tight, the constraint is measured at $z=0$ and therefore inapplicable to probe a cosmological transition occurring in the dark ages.

As such, the constraints that we discussed are in principle not directly applicable to the electron mass bounds above, as long as the near-instantaneous transition is kept outside of the electromagnetically observable range. However, one can wonder how to physically realize such a transition and therefore the study of these constraints will be vital in the context of our arguments on physical models in \cref{sec:physical}.

\subsection{Summarizing the viability of constant shifts of the electron mass}

We recognize that the data does not disfavor constant variations of the electron mass with a near-instantaneous transition in the dark ages. Instead, we find a slight preference for models involving such a transition typically at the $\sim 2\sigma$ level with DESI data. Especially models that feature additional late-time components such as curvature allow for a $\sim 3.6\sigma$ preference, which is robust to supernovae data and is not strongly constrained from BBN, which can only put approximate bounds at the level of $\sim 5\%$. However, beyond this purely phenomenological study, the important question arises if such a model can be realized in a consistent theory or if such a theory brings with it additional constraints that rule out the model.

\section{Physical models: scalar fields and electron mass variation}\label{sec:physical}

It is apparent that an instantaneous transition of the value of the electron mass is hard to realize in any dynamical framework. However, the question of how fast a variation of the electron mass can proceed is of high importance for studying such models. If it can be shown that such a variation can not proceed fast enough, it could exclude variations of the electron mass as viable cosmological models altogether.

A more consistent implementation of varying constants must be done at the Lagrangian level. In our current understanding it is the only viable path in order to preserve consistency of the theory and to keep all the relevant symmetries. The simplest implementations of varying masses involve one or several scalar fields. Depending on the Lagrangian implementation and its motivation, one or several constants can vary in coupled ways, typically one would expect a coupled variation of all constants within a theoretical framework motivated by GUTs or string theories \cite{Campbell1995,Coc:2006sx,Damour1994}.

The simplest phenomenological model one can propose is to promote $m_e/m_{e,0}$ itself -- and thus $\mu$ -- to a scalar field. A canonical Lagrangian $\mathcal{L}=\partial_\mu\partial^\mu\phi/2 - V(\phi)$ can be associated to the field, which enters in the Dirac equation for fermions through $m_e/m_{e,0} = e^{\phi}$. Such a model, proposed by \cite{Barrow:2005}, generalizes the so-called Bekenstein models in which the electron charge $e/e_{0}$ is promoted to a scalar field driving variations of $\alpha$ \cite{bekensteinOriginal,Sandvik2002,Olive2002}.\footnote{Such a model can probably be best understood as one where the underlying Yukawa coupling $h_e$ (see \cref{tab:fundamental}) is varied instead, giving the same overall observational effect.} To our best knowledge, the latest constraints on this model can be found in \cite{Scoccola2008}. A coupled variation of both $\alpha$ and $\mu$ in this framework is discussed in \cite{Avelino2008}. 
From a high energy physics perspective, a more motivated approach is to add some dynamics in the Higgs sector. \cite{Calmet2017} proposed that the cosmological evolution of the Higgs field itself could lead to a dynamics of its vacuum expectation value $v$, leading to a variation of all the particle's masses.\footnote{See also \cite{Fung2021} for a similar effect emerging from the coupling of the Higgs field to an axionic field.} However, as $v$ contributes almost entirely to $m_e$ but not as much to $m_p$ or $m_n$\,, which are dominated by the strong binding energy, such a model can be mostly understood as a varying electron mass model. Recent constraints can be found in \cite{Chakrabarti2021}.
Another option is to introduce a variation of the Yukawa couplings or to couple a new field through Yukawa-like interactions with the fermions fields. This scenario is for example possible through the Randall-Sundrum mechanism where a Goldberger-Wise scalar field couples to the fermions \cite{vonHarling2017}.

More generally, any GUT attempting to unify the gauge forces at high energy or any quantum gravity theory involving extra dimensions as string theory should witness a coupled variation of all the gauge couplings and particle masses. A canonical example of this is the unavoidable presence of the dilaton field in string theory, a scalar partner of the graviton which induces a space-time variation of all the masses and gauge couplings \cite{Damour1994,Chilba2007}. 

Finally, as suggested by \cite{Olive2008}, a field could be coupled to fermions through Yukawa couplings associated to a screening mechanism which makes the masses variations dependent on the local density. As such, these models could completely avoid local measurements performed on earth and allow for a significant shift of the constants during recombination in the low-density universe. The viability of such a model based on a symmetron field as a possible solution to the $H_0$ tension was further discussed in \cite{Solomon2022}. We will however not further discuss such mechanisms in this work.

Overall, most of the proposed models share a similar structure by involving a canonical scalar field with a specific potential and a given function relating the field to the variation of the mass $\Delta m_e/m_{e,0}=B_{m_e}(\phi)$. Witnessing that the variations of $m_e/m_{e,0}$ are necessarily constrained by local data to be small today, one can assume in full generality that $m_e/m_{e,0}\simeq 1 + \zeta (\phi-\phi_0)$, where $\phi_0$ is today's value of the field and $\zeta = \partial_\phi B_{m_e}|_{\phi=\phi_0}$\,. As such, similar arguments on the speed of the dark ages transition as the ones of \cite{Vacher2024} for $\alpha$-varying models can be applied to single field models.

\subsection{Single-field constraints}\label{ssec:single_field}

Assuming that the variation of the electron mass is driven by a coupling to a single scalar field, it is necessarily subject to constraints stemming from the allowed speed of variation for such a scalar field -- essentially the field cannot typically decelerate faster than through Hubble friction except in fine-tuned circumstances. These have been extensively discussed in \cite{Vacher2024}, and we apply the corresponding formalism here, keeping in mind that the same caveats outlined in that work apply identically. The argument proceeds roughly as follows: 
\begin{itemize}
    \item To reach a cosmologically relevant $\Delta \mu/\mu$ at CMB times but to have a vanishing speed $\mathrm{d}\ln \mu/\mathrm{d}z$ today -- as is required from \cref{ssec:me_local} -- the field needs to rapidly decelerate.
    \item The primary mechanism of deceleration is through Hubble drag, which can only proceed as $\dot{\phi} \propto a^{-3}$, and thus $\mathrm{d}\phi/\mathrm{d}\ln a \propto a^{-3}/H$.
    \item Since there are only three decades in redshift (most of which is during matter domination, with $H \propto a^{-3/2}$ approximately), one only finds roughly $3\cdot (3-3/2) = 4.5$ orders of magnitude that the logarithmic field velocity can decrease. A more detailed calculation gives a factor around~${\sim50\,000}$ \cite{Vacher2024}.
    \item If a percent-level impact at recombination is desired (to give cosmological impact, see for example \cref{ssec:me_cmb,ssec:me_bbn}), then this means that the variation of constants today could be at most $10^{-2}/50\,000 \simeq 2\cdot 10^{-7}$.
\end{itemize}
For fine-structure data this kind of constraint results immediately in an almost-no-go-theorem outlined in~\cite{Vacher2024}, since the constraint from atomic clocks is $\mathrm{d}\ln \alpha/\mathrm{d}\ln a = (-2.3\pm 3.5) \cdot 10^{-9}$ (from \cref{fig:me_atomic}),\footnote{Note that the constraint on $\alpha$ remains intact even when marginalizing over $m_e$ or $g$, since it is driven by the constraint from \cite{Filzinger2023}, which doesn't depend on either $\mu$ or $g_p$ (since it features an octopolar vs quadrupolar atomic transition comparison).} excluding cosmologically relevant $\alpha$ variations at the CMB times. 

For variations in the electron mass, the cut is not quite as clear. This is because the constraints on the electron mass alone are at the level $\mathrm{d}\ln m_e/\mathrm{d}\ln a = (-4.5\pm 3.2) \cdot 10^{-7}$, degrading even further to $\mathrm{d}\ln m_e/\mathrm{d}\ln a = (+4.5\pm 5.4) \cdot 10^{-7}$ once marginalization over variations of the proton gyromagnetic ratio $g_p$ is preformed. This is at the same level as the $2 \cdot 10^{-7}$ computed above. As such, in either case the constraint at CMB times through this argument remains at a few times $10^{-2}$, roughly at the limit of current CMB/BBN constraining power. 

For example, the constraint without marginalization over $g_p$ results in a constraint on the electron mass $\left|(\Delta m_e/m_e)_{z_\mathrm{CMB}}\right| \lesssim (2.3 \pm 1.6)\%$, which according to \cref{tab:electron_mass_cmb} is about half to a third of the constraining power of Planck CMB + BAO. As such, these constraints are not yet strong enough to exclude such models (or provide another almost-no-go-theorem), though we expect future atomic clock measurements to be tightly constricting this parameter space (or lead to the detection of an ongoing drift). Already a factor of 10 in constraining power would be enough to provide an almost-no-go-theorem as in \cite{Vacher2024} for variations of the fine structure constant.  
It is yet unclear how much the future atomic clock measurements will be improved, as they face intrinsic noise limitation \cite{Ludlow2015}. However, new concepts have been proposed which might allow to further increase the precision reached  \cite{Robinson2022}.\footnote{Upcoming nuclear clocks measurements are expecting up to a $10^6$ improvement factor on the time drift of $\alpha$ and could also probe time variations of quark masses \cite{Peik2003,Flambaum2006,Fadeev2020}. Unfortunately, these clocks will certainly not be sensitive to variations of $m_e$ as they are based on nuclear transitions. It is clear that with this precision the arguments of \cref{ssec:coupled} and \cite{Vacher2024} would become even more important, such that these indirect constraints through coupling of fine-structure constant and electron mass would dominate the direct ones.}

There are a few minor complications to the story. For example, one could assume that the $1.6\%$ level constraint discussed above would be quite helpful in further restricting variations of the electron mass that involve curvature or neutrino mass variations (see \cref{tab:electron_mass_cmb}), but in this case the presence of curvature can actually cause a small additional deceleration of the logarithmic field speed (as the Hubble parameter will decrease more slowly compared to matter domination). While in most data combinations the curvature is constrained to have a negligible impact, if only supernovae data are used, for example, the constraint can degrade to a slowdown factor of $70\,000-90\,000$ (former: BAO, latter: Pantheon supernovae) between recombination and now, allowing even higher variations at the CMB by a factor of almost 2. However, we expect that at least the constraints presented for the model involving a shift in the electron mass, curvature, and a non-minimal neutrino mass (see \cref{tab:electron_mass_cmb}, a $3\%$ level constraint) could be constrained using this method, though we leave a more detailed investigation of this exact case for future work.

Beyond this, there are the intrinsic caveats to this method, which relate to potentially fine-tuned shapes of the underlying scalar field potential or the field initial conditions, a very non-trivial coupling, or additional screening mechanisms. See \cite{Vacher2024} for a discussion of these method-inherent caveats.

\subsection{Coupled case}\label{ssec:coupled}
We observe that imposing the electron mass shift to be generated at the Lagrangian level through scalar field puts generous constraints on its shift at CMB times. However, this is not the only way in which a more realistic model might impose additional constraints on the variation of the electron mass. As discussed in \cref{sec:physical}, many realistic models of variations of fundamental couplings (based on strings or GUTs) also impose a coupling of the fine-structure constant. For a review on this topic, see \cite{Martins2017review}. In that review, the author follows \cite{Coc:2006sx} and introduces a simple relationship between electron mass and fine-structure variations
\begin{equation}
        \frac{\Delta m_e}{m_e} = \frac{1+S}{2}   \frac{\Delta \alpha}{\alpha_0}~,
\end{equation}
where the $S$ parameter (see \cite[eq.~(21)]{Coc:2006sx}) is different for most theories, and $S\to\infty$ would be the completely uncoupled theory that has been discussed so far. Cases with $S\simeq160$ and $S=0$ have been considered so far \cite{Coc:2006sx,2010JCAP...01..030N,Martins2021}. It is immediately evident that any model with $S \sim \mathcal{O}(1)$ is subject to constraints of the electron mass variation coming from the fine structure constant variation. With the previous discussion of \cref{ssec:single_field} we would expect such models to be generically excluded from having a large cosmological impact through their electron mass variation.

However, here too constraints can be avoided by following the limit of $S \to \infty$, in which case a given electron mass variation would cause infinitesimally small fine structure constant variations (hence recovering the independent case). Already with $S \sim 80$ the constraints from the fine structure constant and the electron mass variation would be equivalent, and for $S \gtrsim 80$ the original electron mass variation constraints would dominate. Very roughly one can approximate the final level of constraint as
\begin{equation}
    \Delta m_e/m_e|_{z_\mathrm{CMB}} \sim \min\left(2\%, \frac{1+S}{2} \cdot 0.05\%\right)~.
\end{equation}
As an illustration, considering the two cases studied in \cite{Martins2021}, GUTs models with $S\sim 160$ as the ones studied in \cite{Coc:2006sx} would still leave a lot of room for electron mass variations to solve the Hubble tension, while strongly coupled string dilaton models discussed in \cite{Damour2002,Chilba2007,2010JCAP...01..030N} with $S\sim 0$ are already too highly constrained to generate strong variations at recombination (with some additional constraints on their parameter space coming from the universality of free fall data as illustrated in \cite{Vacher2023}).

\section{Discussion and Conclusions}
\label{sec:conclusion}

We have summarized and explained the current status of variations of the electron mass in cosmology, with a particular focus on the variation of the electron mass. We have shown that the simplest model of a constant shift of the electron mass results in phenomenologically interesting consequences regarding the sound horizon and the Hubble tension. We find values as high as $H_0 = (71.61\pm1.00)$ km/s/Mpc (see \cref{tab:electron_mass_cmb}), nicely easing the Hubble tension.

We argue that these models generically lead to a somewhat lower value of $\Omega_m$\,. Interestingly, such a lower value of $\Omega_m$ appears to be favored by current BAO data (such as from DESI), see also \cite{Lynch2024}. We find a tantalizing $2\sigma$ preference in the baseline model for Planck+DESI BAO, and $3.6\sigma$ preference when curvature is added (when also adding the PantheonPLUS supernovae sample). A similar preference is also found when including CPL dynamical dark energy in addition to the electron mass variation. %This case could be realized through the existence of a single scalar field inducing responsible for both dark energy and electron mass variations. 
We also find that such variations are typically not strongly constrained from light element abundance measurements as the impact of electron mass variations on BBN is rather minute, confirming earlier indications by \cite{Seto2023}. Finally, constraints from local observations have to be connected to these shifts using more realistic models. In this work we put first constraints on electron mass variations in single-field models and show that they provide only very mild direct constraints and are not yet at a level to give sufficient constraining power to constrain this solution to the Hubble tension. We note that models including screening mechanisms could avoid these bounds altogether.

Overall we summarize that electron mass variations are still an interesting solution to the Hubble tension and especially now with a tantalizing hint from DESI data. We find no evidence that the cosmologically interesting parameter space can be tightly constrained using current observations.

Looking forward, we expect several advancements in observations to allow for more definite conclusions on this matter. First, measurements of the cosmological recombination radiation would provide immediate and clearly distinguishable signals that would allow clear separations of various early time models \cite{Lucca2024,Hart2023}. Second, more precise CMB data are expected shortly from  ACT, SPT, Simon's observatory, CMB-S4 and LiteBIRD \cite{ACT,SPT,SimonsObservatory,LiteBIRD} along with more precise BAO data from DESI and Euclid \cite{DESI,Euclid} and more precise post-recombination measurements \cite{HIRES}. Any of these upcoming measurements is expected to find either evidence for or against cosmologically relevant mass variations, giving a bright future for the mass effect.

\section*{Acknowledgments}

The authors would like to thank Carlos Martins and Jens Chluba for useful discussions during the preparation of this work, as well as Licia Verde for additional inspiration. The authors would also like to thank James Rich and Zach Weiner for comments on the final version of the draft. The authors are very grateful to our colleagues at LMU and SISSA for their help in improving the clarity of the manuscript and LV would like to thank especially Pranjal Ralegankar for fruitful discussions and comments.

NS acknowledges the support of the following Maria de Maetzu fellowship grant: Esta publicaci\'on es parte de la ayuda CEX2019-000918-M, financiada por MCIN/AEI/10.13039/5\\01100011033. NS also acknowledges support by MICINN grant number PID 2022-141125NB-I00.
LV acknowledges partial support by the Italian Space Agency LiteBIRD Project (ASI Grants No. 2020-9-HH.0 and 2016-24-H.1-2018), as well as the RadioForegroundsPlus Project HORIZON-CL4-2023-SPACE-01, GA 101135036. This work was partially done in the framework of the FCT project 2022.04048.PTDC (Phi in the Sky, DOI 10.54499/2022.04048.PTDC).

\bibliography{apssamp}

\providecommand{\href}[2]{#2}\begingroup\raggedright\begin{thebibliography}{100}

\bibitem{Verde2023}
L.~{Verde}, N.~{Sch{\"o}neberg} and H.~{Gil-Mar{\'\i}n}, \emph{{A tale of many
  $H_0$}}, \href{https://doi.org/10.48550/arXiv.2311.13305}{\emph{arXiv
  e-prints} (2023) arXiv:2311.13305}
  [\href{https://arxiv.org/abs/2311.13305}{{\ttfamily 2311.13305}}].

\bibitem{Planck2018VI}
{\scshape Planck} collaboration, \emph{{Planck 2018 results. VI. Cosmological
  parameters}},
  \href{https://doi.org/10.1051/0004-6361/201833910}{\emph{Astron. Astrophys.}
  {\bfseries 641} (2020) A6}
  [\href{https://arxiv.org/abs/1807.06209}{{\ttfamily 1807.06209}}].

\bibitem{Riess2022}
A.G.~{Riess}, W.~{Yuan}, L.M.~{Macri}, D.~{Scolnic}, D.~{Brout}, S.~{Casertano}
  et~al., \emph{{A Comprehensive Measurement of the Local Value of the Hubble
  Constant with 1 km s$^{-1}$ Mpc$^{-1}$ Uncertainty from the Hubble Space
  Telescope and the SH0ES Team}},
  \href{https://doi.org/10.3847/2041-8213/ac5c5b}{\emph{The Astrophysical
  Journal Letters} {\bfseries 934} (2022) L7}
  [\href{https://arxiv.org/abs/2112.04510}{{\ttfamily 2112.04510}}].

\bibitem{DiValentino:2021izs}
E.~Di~Valentino, O.~Mena, S.~Pan, L.~Visinelli, W.~Yang, A.~Melchiorri et~al.,
  \emph{{In the realm of the Hubble tension\textemdash{}a review of
  solutions}}, \href{https://doi.org/10.1088/1361-6382/ac086d}{\emph{Class.
  Quant. Grav.} {\bfseries 38} (2021) 153001}
  [\href{https://arxiv.org/abs/2103.01183}{{\ttfamily 2103.01183}}].

\bibitem{Damour2003}
T.~{Damour}, \emph{{String theory, cosmology and varying constants}},
  \href{https://doi.org/10.1023/A:1022596316014}{\emph{\apss} {\bfseries 283}
  (2003) 445} [\href{https://arxiv.org/abs/gr-qc/0210059}{{\ttfamily
  gr-qc/0210059}}].

\bibitem{Dine2003}
M.~{Dine}, Y.~{Nir}, G.~{Raz} and T.~{Volansky}, \emph{{Time variations in the
  scale of grand unification}},
  \href{https://doi.org/10.1103/PhysRevD.67.015009}{\emph{\prd} {\bfseries 67}
  (2003) 015009} [\href{https://arxiv.org/abs/hep-ph/0209134}{{\ttfamily
  hep-ph/0209134}}].

\bibitem{Dent2004}
T.~{Dent}, \emph{{Varying alpha, thresholds and fermion masses}},
  \href{https://doi.org/10.1016/j.nuclphysb.2003.10.047}{\emph{Nuclear Physics
  B} {\bfseries 677} (2004) 471}
  [\href{https://arxiv.org/abs/hep-ph/0305026}{{\ttfamily hep-ph/0305026}}].

\bibitem{bekensteinOriginal}
J.D.~{Bekenstein}, \emph{{Fine-structure constant: Is it really a constant?}},
  \href{https://doi.org/10.1103/PhysRevD.25.1527}{\emph{\prd} {\bfseries 25}
  (1982) 1527}.

\bibitem{H0olympics}
N.~{Sch{\"o}neberg}, G.F.~{Abell{\'a}n}, A.~{P{\'e}rez S{\'a}nchez},
  S.J.~{Witte}, V.~{Poulin} and J.~{Lesgourgues}, \emph{{The $H_0$ Olympics: A
  fair ranking of proposed models}}, {\emph{arXiv e-prints} (2021)
  arXiv:2107.10291} [\href{https://arxiv.org/abs/2107.10291}{{\ttfamily
  2107.10291}}].

\bibitem{Hart:2021kad}
L.~Hart and J.~Chluba, \emph{{Varying fundamental constants principal component
  analysis: additional hints about the Hubble tension}},
  \href{https://doi.org/10.1093/mnras/stab2777}{\emph{Mon. Not. Roy. Astron.
  Soc.} {\bfseries 510} (2022) 2206}
  [\href{https://arxiv.org/abs/2107.12465}{{\ttfamily 2107.12465}}].

\bibitem{Chluba2023}
J.~{Chluba} and L.~{Hart}, \emph{{Varying fundamental constants meet Hubble}},
  \href{https://doi.org/10.48550/arXiv.2309.12083}{\emph{arXiv e-prints} (2023)
  arXiv:2309.12083} [\href{https://arxiv.org/abs/2309.12083}{{\ttfamily
  2309.12083}}].

\bibitem{Lee2023}
N.~{Lee}, Y.~{Ali-Ha{\"\i}moud}, N.~{Sch{\"o}neberg} and V.~{Poulin},
  \emph{{What It Takes to Solve the Hubble Tension through Modifications of
  Cosmological Recombination}},
  \href{https://doi.org/10.1103/PhysRevLett.130.161003}{\emph{\prl} {\bfseries
  130} (2023) 161003} [\href{https://arxiv.org/abs/2212.04494}{{\ttfamily
  2212.04494}}].

\bibitem{Hart:2019dxi}
L.~Hart and J.~Chluba, \emph{{Updated fundamental constant constraints from
  Planck 2018 data and possible relations to the Hubble tension}},
  \href{https://doi.org/10.1093/mnras/staa412}{\emph{Mon. Not. Roy. Astron.
  Soc.} {\bfseries 493} (2020) 3255}
  [\href{https://arxiv.org/abs/1912.03986}{{\ttfamily 1912.03986}}].

\bibitem{Sekiguchi2021}
T.~{Sekiguchi} and T.~{Takahashi}, \emph{{Early recombination as a solution to
  the H$_{0}$ tension}},
  \href{https://doi.org/10.1103/PhysRevD.103.083507}{\emph{\prd} {\bfseries
  103} (2021) 083507} [\href{https://arxiv.org/abs/2007.03381}{{\ttfamily
  2007.03381}}].

\bibitem{Smith:2018rnu}
T.L.~Smith, D.~Grin, D.~Robinson and D.~Qi, \emph{{Probing spatial variation of
  the fine-structure constant using the CMB}},
  \href{https://doi.org/10.1103/PhysRevD.99.043531}{\emph{Phys. Rev. D}
  {\bfseries 99} (2019) 043531}
  [\href{https://arxiv.org/abs/1808.07486}{{\ttfamily 1808.07486}}].

\bibitem{Will1993}
C.M.~Will, \emph{Theory and Experiment in Gravitational Physics}, Cambridge
  University Press (1993).

\bibitem{Damour2012}
T.~{Damour}, \emph{{Theoretical aspects of the equivalence principle}},
  \href{https://doi.org/10.1088/0264-9381/29/18/184001}{\emph{Classical and
  Quantum Gravity} {\bfseries 29} (2012) 184001}
  [\href{https://arxiv.org/abs/1202.6311}{{\ttfamily 1202.6311}}].

\bibitem{doi:10.1142/S0218271801000822}
M.~Chevallier and D.~Polarski, \emph{Accelerating universes with scaling dark
  matter}, \href{https://doi.org/10.1142/S0218271801000822}{\emph{International
  Journal of Modern Physics D} {\bfseries 10} (2001) 213}
  [\href{https://arxiv.org/abs/https://doi.org/10.1142/S0218271801000822}{{\ttfamily
  https://doi.org/10.1142/S0218271801000822}}].

\bibitem{2003PhRvL..90i1301L}
E.V.~{Linder}, \emph{{Exploring the Expansion History of the Universe}},
  \href{https://doi.org/10.1103/PhysRevLett.90.091301}{\emph{\prl} {\bfseries
  90} (2003) 091301} [\href{https://arxiv.org/abs/astro-ph/0208512}{{\ttfamily
  astro-ph/0208512}}].

\bibitem{Burns:2021pkx}
J.~{Burns}, G.~{Hallinan}, T.-C.~{Chang}, M.~{Anderson}, J.~{Bowman},
  R.~{Bradley} et~al., \emph{{A Lunar Farside Low Radio Frequency Array for
  Dark Ages 21-cm Cosmology}},
  \href{https://doi.org/10.48550/arXiv.2103.08623}{\emph{arXiv e-prints} (2021)
  arXiv:2103.08623} [\href{https://arxiv.org/abs/2103.08623}{{\ttfamily
  2103.08623}}].

\bibitem{2011MNRAS.411..955M}
A.~{Mesinger}, S.~{Furlanetto} and R.~{Cen}, \emph{{21CMFAST: a fast,
  seminumerical simulation of the high-redshift 21-cm signal}},
  \href{https://doi.org/10.1111/j.1365-2966.2010.17731.x}{\emph{\mnras}
  {\bfseries 411} (2011) 955}
  [\href{https://arxiv.org/abs/1003.3878}{{\ttfamily 1003.3878}}].

\bibitem{2019BAAS...51g.241P}
A.~{Parsons}, J.E.~{Aguirre}, A.P.~{Beardsley}, G.~{Bernardi}, J.D.~{Bowman},
  P.~{Bull} et~al., \emph{{A Roadmap for Astrophysics and Cosmology with
  High-Redshift 21 cm Intensity Mapping}},  in \emph{Bulletin of the American
  Astronomical Society}, vol.~51, p.~241, Sept., 2019,
  \href{https://doi.org/10.48550/arXiv.1907.06440}{DOI}
  [\href{https://arxiv.org/abs/1907.06440}{{\ttfamily 1907.06440}}].

\bibitem{DESI:2024uvr}
{\scshape DESI} collaboration, \emph{{DESI 2024 III: Baryon Acoustic
  Oscillations from Galaxies and Quasars}},
  \href{https://arxiv.org/abs/2404.03000}{{\ttfamily 2404.03000}}.

\bibitem{DESI:2024lzq}
{\scshape DESI} collaboration, \emph{{DESI 2024 IV: Baryon Acoustic
  Oscillations from the Lyman Alpha Forest}},
  \href{https://arxiv.org/abs/2404.03001}{{\ttfamily 2404.03001}}.

\bibitem{Scolnic:2021amr}
D.~Scolnic et~al., \emph{{The Pantheon+ Analysis: The Full Data Set and
  Light-curve Release}},
  \href{https://doi.org/10.3847/1538-4357/ac8b7a}{\emph{Astrophys. J.}
  {\bfseries 938} (2022) 113}
  [\href{https://arxiv.org/abs/2112.03863}{{\ttfamily 2112.03863}}].

\bibitem{Hart:2017ndk}
L.~Hart and J.~Chluba, \emph{{New constraints on time-dependent variations of
  fundamental constants using Planck data}},
  \href{https://doi.org/10.1093/mnras/stx2783}{\emph{Mon. Not. Roy. Astron.
  Soc.} {\bfseries 474} (2018) 1850}
  [\href{https://arxiv.org/abs/1705.03925}{{\ttfamily 1705.03925}}].

\bibitem{mukhanov2005physical}
V.~Mukhanov, \emph{Physical Foundations of Cosmology}, Physical Foundations of
  Cosmology, Cambridge University Press (2005).

\bibitem{coles2003cosmology}
P.~Coles and P.~Lucchin, \emph{Cosmology: The Origin and Evolution of Cosmic
  Structure}, Wiley (2003).

\bibitem{weinberg2008cosmology}
S.~Weinberg, \emph{Cosmology}, OUP Oxford (2008).

\bibitem{dodelson2003modern}
S.~Dodelson, \emph{Modern Cosmology}, Elsevier Science (2003).

\bibitem{Seager:1999bc}
S.~Seager, D.D.~Sasselov and D.~Scott, \emph{{A new calculation of the
  recombination epoch}}, \href{https://doi.org/10.1086/312250}{\emph{Astrophys.
  J. Lett.} {\bfseries 523} (1999) L1}
  [\href{https://arxiv.org/abs/astro-ph/9909275}{{\ttfamily
  astro-ph/9909275}}].

\bibitem{Seager:1999km}
S.~Seager, D.D.~Sasselov and D.~Scott, \emph{{How exactly did the universe
  become neutral?}}, \href{https://doi.org/10.1086/313388}{\emph{Astrophys. J.
  Suppl.} {\bfseries 128} (2000) 407}
  [\href{https://arxiv.org/abs/astro-ph/9912182}{{\ttfamily
  astro-ph/9912182}}].

\bibitem{Ali-Haimoud:2010hou}
Y.~Ali-Haimoud and C.M.~Hirata, \emph{{HyRec: A fast and highly accurate
  primordial hydrogen and helium recombination code}},
  \href{https://doi.org/10.1103/PhysRevD.83.043513}{\emph{Phys. Rev. D}
  {\bfseries 83} (2011) 043513}
  [\href{https://arxiv.org/abs/1011.3758}{{\ttfamily 1011.3758}}].

\bibitem{2013ascl.soft04017C}
J.~{Chluba} and R.M.~{Thomas}, ``{CosmoRec: Cosmological Recombination code}.''
  Astrophysics Source Code Library, record ascl:1304.017, Apr., 2013.

\bibitem{Chiang:2018xpn}
C.-T.~Chiang and A.~Slosar, \emph{{Inferences of $H_0$ in presence of a
  non-standard recombination}},
  \href{https://arxiv.org/abs/1811.03624}{{\ttfamily 1811.03624}}.

\bibitem{Baryakhtar2024}
M.~{Baryakhtar}, O.~{Simon} and Z.J.~{Weiner}, \emph{{Varying-constant
  cosmology from hyperlight, coupled scalars}},
  \href{https://doi.org/10.48550/arXiv.2405.10358}{\emph{arXiv e-prints} (2024)
  arXiv:2405.10358} [\href{https://arxiv.org/abs/2405.10358}{{\ttfamily
  2405.10358}}].

\bibitem{1995ApJ...444..489H}
W.~{Hu} and N.~{Sugiyama}, \emph{{Anisotropies in the Cosmic Microwave
  Background: an Analytic Approach}},
  \href{https://doi.org/10.1086/175624}{\emph{\apj} {\bfseries 444} (1995) 489}
  [\href{https://arxiv.org/abs/astro-ph/9407093}{{\ttfamily
  astro-ph/9407093}}].

\bibitem{1995PhRvD..51.2599H}
W.~{Hu} and N.~{Sugiyama}, \emph{{Toward understanding CMB anisotropies and
  their implications}},
  \href{https://doi.org/10.1103/PhysRevD.51.2599}{\emph{\prd} {\bfseries 51}
  (1995) 2599} [\href{https://arxiv.org/abs/astro-ph/9411008}{{\ttfamily
  astro-ph/9411008}}].

\bibitem{1996ApJ...471..542H}
W.~{Hu} and N.~{Sugiyama}, \emph{{Small-Scale Cosmological Perturbations: an
  Analytic Approach}}, \href{https://doi.org/10.1086/177989}{\emph{\apj}
  {\bfseries 471} (1996) 542}
  [\href{https://arxiv.org/abs/astro-ph/9510117}{{\ttfamily
  astro-ph/9510117}}].

\bibitem{2009ApJS..180..330K}
E.~{Komatsu}, J.~{Dunkley}, M.R.~{Nolta}, C.L.~{Bennett}, B.~{Gold},
  G.~{Hinshaw} et~al., \emph{{Five-Year Wilkinson Microwave Anisotropy Probe
  Observations: Cosmological Interpretation}},
  \href{https://doi.org/10.1088/0067-0049/180/2/330}{\emph{\apjs} {\bfseries
  180} (2009) 330} [\href{https://arxiv.org/abs/0803.0547}{{\ttfamily
  0803.0547}}].

\bibitem{Chen:2018dbv}
L.~Chen, Q.-G.~Huang and K.~Wang, \emph{{Distance Priors from Planck Final
  Release}}, \href{https://doi.org/10.1088/1475-7516/2019/02/028}{\emph{JCAP}
  {\bfseries 02} (2019) 028}
  [\href{https://arxiv.org/abs/1808.05724}{{\ttfamily 1808.05724}}].

\bibitem{Kaplinghat:1998ry}
M.~Kaplinghat, R.J.~Scherrer and M.S.~Turner, \emph{{Constraining variations in
  the fine structure constant with the cosmic microwave background}},
  \href{https://doi.org/10.1103/PhysRevD.60.023516}{\emph{Phys. Rev. D}
  {\bfseries 60} (1999) 023516}
  [\href{https://arxiv.org/abs/astro-ph/9810133}{{\ttfamily
  astro-ph/9810133}}].

\bibitem{Avelino:2000ea}
P.P.~Avelino, C.J.A.P.~Martins, G.~Rocha and P.T.P.~Viana, \emph{{Looking for a
  varying alpha in the cosmic microwave background}},
  \href{https://doi.org/10.1103/PhysRevD.62.123508}{\emph{Phys. Rev. D}
  {\bfseries 62} (2000) 123508}
  [\href{https://arxiv.org/abs/astro-ph/0008446}{{\ttfamily
  astro-ph/0008446}}].

\bibitem{Battye:2000ds}
R.A.~Battye, R.~Crittenden and J.~Weller, \emph{{Cosmic concordance and the
  fine structure constant}},
  \href{https://doi.org/10.1103/PhysRevD.63.043505}{\emph{Phys. Rev. D}
  {\bfseries 63} (2001) 043505}
  [\href{https://arxiv.org/abs/astro-ph/0008265}{{\ttfamily
  astro-ph/0008265}}].

\bibitem{Avelino:2001nr}
P.P.~Avelino, S.~Esposito, G.~Mangano, C.J.A.P.~Martins, A.~Melchiorri,
  G.~Miele et~al., \emph{{Early universe constraints on a time varying fine
  structure constant}},
  \href{https://doi.org/10.1103/PhysRevD.64.103505}{\emph{Phys. Rev. D}
  {\bfseries 64} (2001) 103505}
  [\href{https://arxiv.org/abs/astro-ph/0102144}{{\ttfamily
  astro-ph/0102144}}].

\bibitem{Martins:2003pe}
C.J.A.P.~Martins, A.~Melchiorri, G.~Rocha, R.~Trotta, P.P.~Avelino and
  P.T.P.~Viana, \emph{{Wmap constraints on varying alpha and the promise of
  reionization}},
  \href{https://doi.org/10.1016/j.physletb.2003.11.080}{\emph{Phys. Lett. B}
  {\bfseries 585} (2004) 29}
  [\href{https://arxiv.org/abs/astro-ph/0302295}{{\ttfamily
  astro-ph/0302295}}].

\bibitem{Martins:2010gu}
C.J.A.P.~Martins, E.~Menegoni, S.~Galli, G.~Mangano and A.~Melchiorri,
  \emph{{Varying couplings in the early universe: correlated variations of
  $\alpha$ and $G$}},
  \href{https://doi.org/10.1103/PhysRevD.82.023532}{\emph{Phys. Rev. D}
  {\bfseries 82} (2010) 023532}
  [\href{https://arxiv.org/abs/1001.3418}{{\ttfamily 1001.3418}}].

\bibitem{Thiele:2021okz}
L.~Thiele, Y.~Guan, J.C.~Hill, A.~Kosowsky and D.N.~Spergel, \emph{{Can
  small-scale baryon inhomogeneities resolve the Hubble tension? An
  investigation with ACT DR4}},
  \href{https://doi.org/10.1103/PhysRevD.104.063535}{\emph{Phys. Rev. D}
  {\bfseries 104} (2021) 063535}
  [\href{https://arxiv.org/abs/2105.03003}{{\ttfamily 2105.03003}}].

\bibitem{Galli:2021mxk}
S.~Galli, L.~Pogosian, K.~Jedamzik and L.~Balkenhol, \emph{{Consistency of
  Planck, ACT, and SPT constraints on magnetically assisted recombination and
  forecasts for future experiments}},
  \href{https://doi.org/10.1103/PhysRevD.105.023513}{\emph{Phys. Rev. D}
  {\bfseries 105} (2022) 023513}
  [\href{https://arxiv.org/abs/2109.03816}{{\ttfamily 2109.03816}}].

\bibitem{Jedamzik:2023rfd}
K.~Jedamzik, T.~Abel and Y.~Ali-Haimoud, \emph{{Cosmic Recombination in the
  Presence of Primordial Magnetic Fields}},
  \href{https://arxiv.org/abs/2312.11448}{{\ttfamily 2312.11448}}.

\bibitem{Lynch2024}
G.P.~{Lynch}, L.~{Knox} and J.~{Chluba}, \emph{{DESI and the Hubble tension in
  light of modified recombination}},
  \href{https://doi.org/10.48550/arXiv.2406.10202}{\emph{arXiv e-prints} (2024)
  arXiv:2406.10202} [\href{https://arxiv.org/abs/2406.10202}{{\ttfamily
  2406.10202}}].

\bibitem{DESI:2024mwx}
{DESI Collaboration}, A.G.~{Adame}, J.~{Aguilar}, S.~{Ahlen}, S.~{Alam},
  D.M.~{Alexander} et~al., \emph{{DESI 2024 VI: Cosmological Constraints from
  the Measurements of Baryon Acoustic Oscillations}},
  \href{https://doi.org/10.48550/arXiv.2404.03002}{\emph{arXiv e-prints} (2024)
  arXiv:2404.03002} [\href{https://arxiv.org/abs/2404.03002}{{\ttfamily
  2404.03002}}].

\bibitem{Khalife2024}
A.R.~{Khalife}, M.B.~{Zanjani}, S.~{Galli}, S.~{G{\"u}nther}, J.~{Lesgourgues}
  and K.~{Benabed}, \emph{{Review of Hubble tension solutions with new SH0ES
  and SPT-3G data}},
  \href{https://doi.org/10.1088/1475-7516/2024/04/059}{\emph{\jcap} {\bfseries
  2024} (2024) 059} [\href{https://arxiv.org/abs/2312.09814}{{\ttfamily
  2312.09814}}].

\bibitem{2011JCAP...07..034B}
D.~{Blas}, J.~{Lesgourgues} and T.~{Tram}, \emph{{The Cosmic Linear Anisotropy
  Solving System (CLASS). Part II: Approximation schemes}},
  \href{https://doi.org/10.1088/1475-7516/2011/07/034}{\emph{\jcap} {\bfseries
  2011} (2011) 034} [\href{https://arxiv.org/abs/1104.2933}{{\ttfamily
  1104.2933}}].

\bibitem{Audren:2012wb}
B.~Audren, J.~Lesgourgues, K.~Benabed and S.~Prunet, \emph{{Conservative
  Constraints on Early Cosmology: an illustration of the Monte Python
  cosmological parameter inference code}},
  \href{https://doi.org/10.1088/1475-7516/2013/02/001}{\emph{JCAP} {\bfseries
  1302} (2013) 001} [\href{https://arxiv.org/abs/1210.7183}{{\ttfamily
  1210.7183}}].

\bibitem{Brinckmann:2018}
T.~{Brinckmann} and J.~{Lesgourgues}, \emph{{MontePython 3: Boosted MCMC
  sampler and other features}},
  \href{https://doi.org/10.1016/j.dark.2018.100260}{\emph{Physics of the Dark
  Universe} {\bfseries 24} (2019) 100260}
  [\href{https://arxiv.org/abs/1804.07261}{{\ttfamily 1804.07261}}].

\bibitem{Lewis:2019xzd}
A.~Lewis, \emph{{GetDist: a Python package for analysing Monte Carlo samples}},
   \href{https://arxiv.org/abs/1910.13970}{{\ttfamily 1910.13970}}.

\bibitem{Riess2021}
A.G.~{Riess}, S.~{Casertano}, W.~{Yuan}, J.B.~{Bowers}, L.~{Macri}, J.C.~{Zinn}
  et~al., \emph{{Cosmic Distances Calibrated to 1\% Precision with Gaia EDR3
  Parallaxes and Hubble Space Telescope Photometry of 75 Milky Way Cepheids
  Confirm Tension with {\ensuremath{\Lambda}}CDM}},
  \href{https://doi.org/10.3847/2041-8213/abdbaf}{\emph{The Astrophysical
  Journal Letters} {\bfseries 908} (2021) L6}
  [\href{https://arxiv.org/abs/2012.08534}{{\ttfamily 2012.08534}}].

\bibitem{Toda:2024ncp}
Y.~Toda, W.~Giar\`e, E.~\"Oz\"ulker, E.~Di~Valentino and S.~Vagnozzi,
  \emph{{Combining pre- and post-recombination new physics to address
  cosmological tensions: case study with varying electron mass and a
  sign-switching cosmological constant}},
  \href{https://arxiv.org/abs/2407.01173}{{\ttfamily 2407.01173}}.

\bibitem{Seto2024}
O.~{Seto} and Y.~{Toda}, \emph{{DESI constraints on varying electron mass model
  and axion-like early dark energy}},
  \href{https://doi.org/10.48550/arXiv.2405.11869}{\emph{arXiv e-prints} (2024)
  arXiv:2405.11869} [\href{https://arxiv.org/abs/2405.11869}{{\ttfamily
  2405.11869}}].

\bibitem{Planck2015XIV}
{\scshape Planck} collaboration, \emph{{Planck 2015 results. XIV. Dark energy
  and modified gravity}},
  \href{https://doi.org/10.1051/0004-6361/201525814}{\emph{Astron. Astrophys.}
  {\bfseries 594} (2016) A14}
  [\href{https://arxiv.org/abs/1502.01590}{{\ttfamily 1502.01590}}].

\bibitem{Planck2018V}
{Planck Collaboration}, N.~{Aghanim}, Y.~{Akrami}, M.~{Ashdown}, J.~{Aumont},
  C.~{Baccigalupi} et~al., \emph{{Planck 2018 results. V. CMB power spectra and
  likelihoods}}, \href{https://doi.org/10.1051/0004-6361/201936386}{\emph{\aap}
  {\bfseries 641} (2020) A5}
  [\href{https://arxiv.org/abs/1907.12875}{{\ttfamily 1907.12875}}].

\bibitem{BOSS:2016wmc}
{\scshape BOSS} collaboration, \emph{{The clustering of galaxies in the
  completed SDSS-III Baryon Oscillation Spectroscopic Survey: cosmological
  analysis of the DR12 galaxy sample}},
  \href{https://doi.org/10.1093/mnras/stx721}{\emph{Mon. Not. Roy. Astron.
  Soc.} {\bfseries 470} (2017) 2617}
  [\href{https://arxiv.org/abs/1607.03155}{{\ttfamily 1607.03155}}].

\bibitem{eBOSS:2019qwo}
{\scshape eBOSS} collaboration, \emph{{Baryon acoustic oscillations from the
  cross-correlation of Ly$\alpha$ absorption and quasars in eBOSS DR14}},
  \href{https://doi.org/10.1051/0004-6361/201935641}{\emph{Astron. Astrophys.}
  {\bfseries 629} (2019) A86}
  [\href{https://arxiv.org/abs/1904.03430}{{\ttfamily 1904.03430}}].

\bibitem{eBOSS:2019ytm}
{\scshape eBOSS} collaboration, \emph{{Baryon acoustic oscillations at z = 2.34
  from the correlations of Ly$\alpha$ absorption in eBOSS DR14}},
  \href{https://doi.org/10.1051/0004-6361/201935638}{\emph{Astron. Astrophys.}
  {\bfseries 629} (2019) A85}
  [\href{https://arxiv.org/abs/1904.03400}{{\ttfamily 1904.03400}}].

\bibitem{Ross:2014qpa}
A.J.~Ross, L.~Samushia, C.~Howlett, W.J.~Percival, A.~Burden and M.~Manera,
  \emph{{The clustering of the SDSS DR7 main Galaxy sample \textendash{} I. A 4
  per cent distance measure at $z = 0.15$}},
  \href{https://doi.org/10.1093/mnras/stv154}{\emph{Mon. Not. Roy. Astron.
  Soc.} {\bfseries 449} (2015) 835}
  [\href{https://arxiv.org/abs/1409.3242}{{\ttfamily 1409.3242}}].

\bibitem{2011MNRAS.416.3017B}
F.~{Beutler}, C.~{Blake}, M.~{Colless}, D.H.~{Jones}, L.~{Staveley-Smith},
  L.~{Campbell} et~al., \emph{{The 6dF Galaxy Survey: baryon acoustic
  oscillations and the local Hubble constant}},
  \href{https://doi.org/10.1111/j.1365-2966.2011.19250.x}{\emph{\mnras}
  {\bfseries 416} (2011) 3017}
  [\href{https://arxiv.org/abs/1106.3366}{{\ttfamily 1106.3366}}].

\bibitem{2018ApJ...859..101S}
D.M.~{Scolnic}, D.O.~{Jones}, A.~{Rest}, Y.C.~{Pan}, R.~{Chornock},
  R.J.~{Foley} et~al., \emph{{The Complete Light-curve Sample of
  Spectroscopically Confirmed SNe Ia from Pan-STARRS1 and Cosmological
  Constraints from the Combined Pantheon Sample}},
  \href{https://doi.org/10.3847/1538-4357/aab9bb}{\emph{\apj} {\bfseries 859}
  (2018) 101} [\href{https://arxiv.org/abs/1710.00845}{{\ttfamily
  1710.00845}}].

\bibitem{2021ApJ...919...16F}
W.L.~{Freedman}, \emph{{Measurements of the Hubble Constant: Tensions in
  Perspective}}, \href{https://doi.org/10.3847/1538-4357/ac0e95}{\emph{\apj}
  {\bfseries 919} (2021) 16}
  [\href{https://arxiv.org/abs/2106.15656}{{\ttfamily 2106.15656}}].

\bibitem{Iocco:2008va}
F.~Iocco, G.~Mangano, G.~Miele, O.~Pisanti and P.D.~Serpico, \emph{{Primordial
  Nucleosynthesis: from precision cosmology to fundamental physics}},
  \href{https://doi.org/10.1016/j.physrep.2009.02.002}{\emph{Phys. Rept.}
  {\bfseries 472} (2009) 1} [\href{https://arxiv.org/abs/0809.0631}{{\ttfamily
  0809.0631}}].

\bibitem{Fields2023}
B.D.~Fields, \emph{Big bang nucleosynthesis: Nuclear physics in the early
  universe},  in \emph{Handbook of Nuclear Physics}, I.~Tanihata, H.~Toki and
  T.~Kajino, eds., (Singapore), pp.~3379--3405, Springer Nature Singapore
  (2023), \href{https://doi.org/10.1007/978-981-19-6345-2_111}{DOI}.

\bibitem{Cyburt:2015mya}
R.H.~Cyburt, B.D.~Fields, K.A.~Olive and T.-H.~Yeh, \emph{{Big Bang
  Nucleosynthesis: 2015}},
  \href{https://doi.org/10.1103/RevModPhys.88.015004}{\emph{Rev. Mod. Phys.}
  {\bfseries 88} (2016) 015004}
  [\href{https://arxiv.org/abs/1505.01076}{{\ttfamily 1505.01076}}].

\bibitem{Burns:2023sgx}
A.-K.~Burns, T.M.P.~Tait and M.~Valli, \emph{{PRyMordial: the first three
  minutes, within and beyond the standard model}},
  \href{https://doi.org/10.1140/epjc/s10052-024-12442-0}{\emph{Eur. Phys. J. C}
  {\bfseries 84} (2024) 86} [\href{https://arxiv.org/abs/2307.07061}{{\ttfamily
  2307.07061}}].

\bibitem{Seto2023}
O.~{Seto} and Y.~{Toda}, \emph{{Big bang nucleosynthesis constraints on varying
  electron mass solution to the Hubble tension}},
  \href{https://doi.org/10.1103/PhysRevD.107.083512}{\emph{\prd} {\bfseries
  107} (2023) 083512} [\href{https://arxiv.org/abs/2206.13209}{{\ttfamily
  2206.13209}}].

\bibitem{2022CoPhC.27108205G}
S.~{Gariazzo}, P.~{F. de Salas}, O.~{Pisanti} and R.~{Consiglio},
  \emph{{PArthENoPE revolutions}},
  \href{https://doi.org/10.1016/j.cpc.2021.108205}{\emph{Computer Physics
  Communications} {\bfseries 271} (2022) 108205}
  [\href{https://arxiv.org/abs/2103.05027}{{\ttfamily 2103.05027}}].

\bibitem{Schoneberg:2024ifp}
N.~Sch\"oneberg, \emph{{The 2024 BBN baryon abundance update}},
  \href{https://doi.org/10.1088/1475-7516/2024/06/006}{\emph{JCAP} {\bfseries
  06} (2024) 006} [\href{https://arxiv.org/abs/2401.15054}{{\ttfamily
  2401.15054}}].

\bibitem{ParticleDataGroup:2022pth}
{\scshape Particle Data Group} collaboration, \emph{{Review of Particle
  Physics}}, \href{https://doi.org/10.1093/ptep/ptac097}{\emph{PTEP} {\bfseries
  2022} (2022) 083C01}.

\bibitem{Schoneberg:2019wmt}
N.~Sch\"oneberg, J.~Lesgourgues and D.C.~Hooper, \emph{{The BAO+BBN take on the
  Hubble tension}},
  \href{https://doi.org/10.1088/1475-7516/2019/10/029}{\emph{JCAP} {\bfseries
  10} (2019) 029} [\href{https://arxiv.org/abs/1907.11594}{{\ttfamily
  1907.11594}}].

\bibitem{Schoneberg:2022ggi}
N.~Sch\"oneberg, L.~Verde, H.~Gil-Mar\'\i{}n and S.~Brieden, \emph{{BAO+BBN
  revisited \textemdash{} growing the Hubble tension with a 0.7 km/s/Mpc
  constraint}},
  \href{https://doi.org/10.1088/1475-7516/2022/11/039}{\emph{JCAP} {\bfseries
  11} (2022) 039} [\href{https://arxiv.org/abs/2209.14330}{{\ttfamily
  2209.14330}}].

\bibitem{Addison:2013haa}
G.E.~Addison, G.~Hinshaw and M.~Halpern, \emph{{Cosmological constraints from
  baryon acoustic oscillations and clustering of large-scale structure}},
  \href{https://doi.org/10.1093/mnras/stt1687}{\emph{Mon. Not. Roy. Astron.
  Soc.} {\bfseries 436} (2013) 1674}
  [\href{https://arxiv.org/abs/1304.6984}{{\ttfamily 1304.6984}}].

\bibitem{BOSS:2014hhw}
{\scshape BOSS} collaboration, \emph{{Cosmological implications of baryon
  acoustic oscillation measurements}},
  \href{https://doi.org/10.1103/PhysRevD.92.123516}{\emph{Phys. Rev. D}
  {\bfseries 92} (2015) 123516}
  [\href{https://arxiv.org/abs/1411.1074}{{\ttfamily 1411.1074}}].

\bibitem{Addison:2017fdm}
G.E.~Addison, D.J.~Watts, C.L.~Bennett, M.~Halpern, G.~Hinshaw and
  J.L.~Weiland, \emph{{Elucidating $\Lambda$CDM: Impact of Baryon Acoustic
  Oscillation Measurements on the Hubble Constant Discrepancy}},
  \href{https://doi.org/10.3847/1538-4357/aaa1ed}{\emph{Astrophys. J.}
  {\bfseries 853} (2018) 119}
  [\href{https://arxiv.org/abs/1707.06547}{{\ttfamily 1707.06547}}].

\bibitem{Cuceu:2019for}
A.~Cuceu, J.~Farr, P.~Lemos and A.~Font-Ribera, \emph{{Baryon Acoustic
  Oscillations and the Hubble Constant: Past, Present and Future}},
  \href{https://doi.org/10.1088/1475-7516/2019/10/044}{\emph{JCAP} {\bfseries
  10} (2019) 044} [\href{https://arxiv.org/abs/1906.11628}{{\ttfamily
  1906.11628}}].

\bibitem{Schoneberg:2021qvd}
N.~Sch\"oneberg, G.~Franco~Abell\'an, A.~P\'erez~S\'anchez, S.J.~Witte,
  V.~Poulin and J.~Lesgourgues, \emph{{The H0 Olympics: A fair ranking of
  proposed models}},
  \href{https://doi.org/10.1016/j.physrep.2022.07.001}{\emph{Phys. Rept.}
  {\bfseries 984} (2022) 1} [\href{https://arxiv.org/abs/2107.10291}{{\ttfamily
  2107.10291}}].

\bibitem{Jansen:2013paa}
P.~Jansen, H.L.~Bethlem and W.~Ubachs, \emph{{Perspective: Tipping the scales -
  search for drifting constants from molecular spectra}},
  \href{https://doi.org/10.1063/1.4853735}{\emph{J. Chem. Phys.} {\bfseries
  140} (2014) 010901} [\href{https://arxiv.org/abs/1312.1875}{{\ttfamily
  1312.1875}}].

\bibitem{Ubachs2007}
W.~{Ubachs}, R.~{Buning}, K.S.E.~{Eikema} and E.~{Reinhold}, \emph{{On a
  possible variation of the proton-to-electron mass ratio: H$_{2}$ spectra in
  the line of sight of high-redshift quasars and in the laboratory}},
  \href{https://doi.org/10.1016/j.jms.2006.12.004}{\emph{Journal of Molecular
  Spectroscopy} {\bfseries 241} (2007) 155}.

\bibitem{Uzan2011}
J.-P.~{Uzan}, \emph{{Varying Constants, Gravitation and Cosmology}},
  \href{https://doi.org/10.12942/lrr-2011-2}{\emph{Living Reviews in
  Relativity} {\bfseries 14} (2011) 2}
  [\href{https://arxiv.org/abs/1009.5514}{{\ttfamily 1009.5514}}].

\bibitem{Martins2017review}
C.J.A.P.~{Martins}, \emph{{The status of varying constants: a review of the
  physics, searches and implications}},
  \href{https://doi.org/10.1088/1361-6633/aa860e}{\emph{Reports on Progress in
  Physics} {\bfseries 80} (2017) 126902}
  [\href{https://arxiv.org/abs/1709.02923}{{\ttfamily 1709.02923}}].

\bibitem{Karshenboim:2000rg}
S.G.~Karshenboim, \emph{{Some possibilities for laboratory searches for
  variations of fundamental constants}},
  \href{https://doi.org/10.1139/cjp-78-7-639}{\emph{Can. J. Phys.} {\bfseries
  78} (2000) 639} [\href{https://arxiv.org/abs/physics/0008051}{{\ttfamily
  physics/0008051}}].

\bibitem{doi:10.1126/science.1154622}
T.~Rosenband, D.B.~Hume, P.O.~Schmidt, C.W.~Chou, A.~Brusch, L.~Lorini et~al.,
  \emph{Frequency ratio of {A}l$^+$ and {H}g$^+$ single-ion optical clocks;
  metrology at the 17th decimal place},
  \href{https://doi.org/10.1126/science.1154622}{\emph{Science} {\bfseries 319}
  (2008) 1808}
  [\href{https://arxiv.org/abs/https://www.science.org/doi/pdf/10.1126/science.1154622}{{\ttfamily
  https://www.science.org/doi/pdf/10.1126/science.1154622}}].

\bibitem{2013PhRvL.111f0801L}
N.~{Leefer}, C.T.M.~{Weber}, A.~{Cing{\"o}z}, J.R.~{Torgerson} and D.~{Budker},
  \emph{{New Limits on Variation of the Fine-Structure Constant Using Atomic
  Dysprosium}},
  \href{https://doi.org/10.1103/PhysRevLett.111.060801}{\emph{\prl} {\bfseries
  111} (2013) 060801} [\href{https://arxiv.org/abs/1304.6940}{{\ttfamily
  1304.6940}}].

\bibitem{Shelkovnikov:2008rv}
A.~Shelkovnikov, R.J.~Butcher, C.~Chardonnet and A.~Amy-Klein, \emph{{Stability
  of the proton-to-electron mass ratio}},
  \href{https://doi.org/10.1103/PhysRevLett.100.150801}{\emph{Phys. Rev. Lett.}
  {\bfseries 100} (2008) 150801}
  [\href{https://arxiv.org/abs/0803.1829}{{\ttfamily 0803.1829}}].

\bibitem{2004LNP...648..209F}
M.~{Fischer}, N.~{Kolachevsky}, M.~{Zimmermann}, R.~{Holzwarth}, T.~{Udem},
  T.W.~{Hansch} et~al., \emph{{Precision Spectroscopy of Atomic Hydrogen and
  Variations of Fundamental Constants}},  in \emph{Astrophysics, Clocks and
  Fundamental Constants}, S.G.~{Karshenboim} and E.~{Peik}, eds., vol.~648,
  pp.~209--227 (2004),
  \href{https://doi.org/10.1007/978-3-540-40991-5_13}{DOI}.

\bibitem{2015CRPhy..16..461A}
M.~{Abgrall}, B.~{Chupin}, L.~{De Sarlo}, J.~{Gu{\'e}na}, P.~{Laurent}, Y.~{Le
  Coq} et~al., \emph{{Atomic fountains and optical clocks at SYRTE: Status and
  perspectives}},
  \href{https://doi.org/10.1016/j.crhy.2015.03.010}{\emph{Comptes Rendus
  Physique} {\bfseries 16} (2015) 461}
  [\href{https://arxiv.org/abs/1507.04623}{{\ttfamily 1507.04623}}].

\bibitem{Fortier:2007jf}
T.M.~Fortier et~al., \emph{{Precision atomic spectroscopy for improved limits
  on variation of the fine structure constant and local position invariance}},
  \href{https://doi.org/10.1103/PhysRevLett.98.070801}{\emph{Phys. Rev. Lett.}
  {\bfseries 98} (2007) 070801}.

\bibitem{2014PhRvA..89b3820T}
C.~{Tamm}, N.~{Huntemann}, B.~{Lipphardt}, V.~{Gerginov}, N.~{Nemitz},
  M.~{Kazda} et~al., \emph{{Cs-based optical frequency measurement using
  cross-linked optical and microwave oscillators}},
  \href{https://doi.org/10.1103/PhysRevA.89.023820}{\emph{\pra} {\bfseries 89}
  (2014) 023820} [\href{https://arxiv.org/abs/1310.8190}{{\ttfamily
  1310.8190}}].

\bibitem{2014PhRvL.113u0802H}
N.~{Huntemann}, B.~{Lipphardt}, C.~{Tamm}, V.~{Gerginov}, S.~{Weyers} and
  E.~{Peik}, \emph{{Improved Limit on a Temporal Variation of m$_{p}$/m$_{e}$
  from Comparisons of Yb$^{+}$ and Cs Atomic Clocks}},
  \href{https://doi.org/10.1103/PhysRevLett.113.210802}{\emph{\prl} {\bfseries
  113} (2014) 210802} [\href{https://arxiv.org/abs/1407.4408}{{\ttfamily
  1407.4408}}].

\bibitem{Guena:2017lbc}
J.~Gu\'ena et~al., \emph{{First international comparison of fountain primary
  frequency standards via a long distance optical fiber link}},
  \href{https://doi.org/10.1088/1681-7575/aa65fe}{\emph{Metrologia} {\bfseries
  54} (2017) 348} [\href{https://arxiv.org/abs/1703.02892}{{\ttfamily
  1703.02892}}].

\bibitem{Lange:2020cul}
R.~Lange, N.~Huntemann, J.M.~Rahm, C.~Sanner, H.~Shao, B.~Lipphardt et~al.,
  \emph{{Improved limits for violations of local position invariance from
  atomic clock comparisons}},
  \href{https://doi.org/10.1103/PhysRevLett.126.011102}{\emph{Phys. Rev. Lett.}
  {\bfseries 126} (2021) 011102}
  [\href{https://arxiv.org/abs/2010.06620}{{\ttfamily 2010.06620}}].

\bibitem{Filzinger:2023zrs}
M.~Filzinger, S.~D\"orscher, R.~Lange, J.~Klose, M.~Steinel, E.~Benkler et~al.,
  \emph{{Improved Limits on the Coupling of Ultralight Bosonic Dark Matter to
  Photons from Optical Atomic Clock Comparisons}},
  \href{https://doi.org/10.1103/PhysRevLett.130.253001}{\emph{Phys. Rev. Lett.}
  {\bfseries 130} (2023) 253001}
  [\href{https://arxiv.org/abs/2301.03433}{{\ttfamily 2301.03433}}].

\bibitem{Campbell1995}
B.A.~{Campbell} and K.A.~{Olive}, \emph{{Nucleosynthesis and the time
  dependence of fundamental couplings}},
  \href{https://doi.org/10.1016/0370-2693(94)01652-S}{\emph{Physics Letters B}
  {\bfseries 345} (1995) 429}
  [\href{https://arxiv.org/abs/hep-ph/9411272}{{\ttfamily hep-ph/9411272}}].

\bibitem{Coc:2006sx}
A.~Coc, N.J.~Nunes, K.A.~Olive, J.-P.~Uzan and E.~Vangioni, \emph{{Coupled
  Variations of Fundamental Couplings and Primordial Nucleosynthesis}},
  \href{https://doi.org/10.1103/PhysRevD.76.023511}{\emph{Phys. Rev. D}
  {\bfseries 76} (2007) 023511}
  [\href{https://arxiv.org/abs/astro-ph/0610733}{{\ttfamily
  astro-ph/0610733}}].

\bibitem{Damour1994}
T.~{Damour} and A.M.~{Polyakov}, \emph{{The string dilation and a least
  coupling principle}},
  \href{https://doi.org/10.1016/0550-3213(94)90143-0}{\emph{Nuclear Physics B}
  {\bfseries 423} (1994) 532}
  [\href{https://arxiv.org/abs/hep-th/9401069}{{\ttfamily hep-th/9401069}}].

\bibitem{Barrow:2005}
J.D.~Barrow and J.~Magueijo, \emph{{Cosmological constraints on a dynamical
  electron mass}},
  \href{https://doi.org/10.1103/PhysRevD.72.043521}{\emph{Phys. Rev. D}
  {\bfseries 72} (2005) 043521}
  [\href{https://arxiv.org/abs/astro-ph/0503222}{{\ttfamily
  astro-ph/0503222}}].

\bibitem{Sandvik2002}
H.B.~{Sandvik}, J.D.~{Barrow} and J.~{Magueijo}, \emph{{A Simple Cosmology with
  a Varying Fine Structure Constant}},
  \href{https://doi.org/10.1103/PhysRevLett.88.031302}{\emph{\prl} {\bfseries
  88} (2002) 031302} [\href{https://arxiv.org/abs/astro-ph/0107512}{{\ttfamily
  astro-ph/0107512}}].

\bibitem{Olive2002}
K.A.~{Olive} and M.~{Pospelov}, \emph{{Evolution of the fine structure constant
  driven by dark matter and the cosmological constant}},
  \href{https://doi.org/10.1103/PhysRevD.65.085044}{\emph{\prd} {\bfseries 65}
  (2002) 085044} [\href{https://arxiv.org/abs/hep-ph/0110377}{{\ttfamily
  hep-ph/0110377}}].

\bibitem{Scoccola2008}
C.G.~Scóccola, M.E.~Mosquera, S.J.~Landau and H.~Vucetich, \emph{Time
  variation of the electron mass in the early universe and the barrow-magueijo
  model}, \href{https://doi.org/10.1086/588086}{\emph{The Astrophysical
  Journal} {\bfseries 681} (2008) 737}.

\bibitem{Avelino2008}
P.P.~{Avelino}, \emph{{Cosmological evolution of {\ensuremath{\alpha}} and
  {\ensuremath{\mu}} and the dynamics of dark energy}},
  \href{https://doi.org/10.1103/PhysRevD.78.043516}{\emph{\prd} {\bfseries 78}
  (2008) 043516} [\href{https://arxiv.org/abs/0804.3394}{{\ttfamily
  0804.3394}}].

\bibitem{Calmet2017}
X.~{Calmet}, \emph{{Cosmological evolution of the Higgs boson's vacuum
  expectation value}},
  \href{https://doi.org/10.1140/epjc/s10052-017-5324-5}{\emph{European Physical
  Journal C} {\bfseries 77} (2017) 729}
  [\href{https://arxiv.org/abs/1707.06922}{{\ttfamily 1707.06922}}].

\bibitem{Fung2021}
L.W.H.~{Fung}, L.~{Li}, T.~{Liu}, H.N.~{Luu}, Y.-C.~{Qiu} and S.H.H.~{Tye},
  \emph{{Axi-Higgs cosmology}},
  \href{https://doi.org/10.1088/1475-7516/2021/08/057}{\emph{\jcap} {\bfseries
  2021} (2021) 057} [\href{https://arxiv.org/abs/2102.11257}{{\ttfamily
  2102.11257}}].

\bibitem{Chakrabarti2021}
S.~{Chakrabarti}, \emph{{Cosmic variation of proton-to-electron mass ratio with
  an interacting Higgs scalar field}},
  \href{https://doi.org/10.1093/mnras/stab1910}{\emph{\mnras} {\bfseries 506}
  (2021) 2518} [\href{https://arxiv.org/abs/2107.00543}{{\ttfamily
  2107.00543}}].

\bibitem{vonHarling2017}
B.~{von Harling} and G.~{Servant}, \emph{{Cosmological evolution of Yukawa
  couplings: the 5D perspective}},
  \href{https://doi.org/10.1007/JHEP05(2017)077}{\emph{Journal of High Energy
  Physics} {\bfseries 2017} (2017) 77}
  [\href{https://arxiv.org/abs/1612.02447}{{\ttfamily 1612.02447}}].

\bibitem{Chilba2007}
T.~{Chiba}, T.~{Kobayashi}, M.~{Yamaguchi} and J.~{Yokoyama}, \emph{{Time
  variation of the proton-electron mass ratio and the fine structure constant
  with a runaway dilaton}},
  \href{https://doi.org/10.1103/PhysRevD.75.043516}{\emph{\prd} {\bfseries 75}
  (2007) 043516} [\href{https://arxiv.org/abs/hep-ph/0610027}{{\ttfamily
  hep-ph/0610027}}].

\bibitem{Olive2008}
K.A.~{Olive} and M.~{Pospelov}, \emph{{Environmental dependence of masses and
  coupling constants}},
  \href{https://doi.org/10.1103/PhysRevD.77.043524}{\emph{\prd} {\bfseries 77}
  (2008) 043524} [\href{https://arxiv.org/abs/0709.3825}{{\ttfamily
  0709.3825}}].

\bibitem{Solomon2022}
R.~{Solomon}, G.~{Agarwal} and D.~{Stojkovic}, \emph{{Environment dependent
  electron mass and the Hubble constant tension}},
  \href{https://doi.org/10.1103/PhysRevD.105.103536}{\emph{\prd} {\bfseries
  105} (2022) 103536} [\href{https://arxiv.org/abs/2201.03127}{{\ttfamily
  2201.03127}}].

\bibitem{Vacher2024}
L.~{Vacher} and N.~{Sch{\"o}neberg}, \emph{{Incompatibility of fine-structure
  constant variations at recombination with local observations}},
  \href{https://doi.org/10.1103/PhysRevD.109.103520}{\emph{\prd} {\bfseries
  109} (2024) 103520} [\href{https://arxiv.org/abs/2403.02256}{{\ttfamily
  2403.02256}}].

\bibitem{Filzinger2023}
M.~{Filzinger}, S.~{D{\"o}rscher}, R.~{Lange}, J.~{Klose}, M.~{Steinel},
  E.~{Benkler} et~al., \emph{{Improved limits on the coupling of ultralight
  bosonic dark matter to photons from optical atomic clock comparisons}},
  \href{https://doi.org/10.48550/arXiv.2301.03433}{\emph{arXiv e-prints} (2023)
  arXiv:2301.03433} [\href{https://arxiv.org/abs/2301.03433}{{\ttfamily
  2301.03433}}].

\bibitem{Ludlow2015}
A.D.~{Ludlow}, M.M.~{Boyd}, J.~{Ye}, E.~{Peik} and P.O.~{Schmidt},
  \emph{{Optical atomic clocks}},
  \href{https://doi.org/10.1103/RevModPhys.87.637}{\emph{Reviews of Modern
  Physics} {\bfseries 87} (2015) 637}
  [\href{https://arxiv.org/abs/1407.3493}{{\ttfamily 1407.3493}}].

\bibitem{Robinson2022}
J.M.~Robinson, M.~Miklos, Y.M.~Tso, C.J.~Kennedy, T.~Bothwell, D.~Kedar et~al.,
  \emph{{Direct comparison of two spin-squeezed optical clock ensembles at the
  10$^{-17}$ level}},
  \href{https://doi.org/10.1038/s41567-023-02310-1}{\emph{Nature Phys.}
  {\bfseries 20} (2024) 208}
  [\href{https://arxiv.org/abs/2211.08621}{{\ttfamily 2211.08621}}].

\bibitem{Peik2003}
E.~Peik and C.~Tamm, \emph{Nuclear laser spectroscopy of the 3.5 ev transition
  in th-229},
  \href{https://doi.org/10.1209/epl/i2003-00210-x}{\emph{Europhysics Letters}
  {\bfseries 61} (2003) 181}.

\bibitem{Flambaum2006}
V.V.~Flambaum, \emph{{Enhanced effect of temporal variation of the fine
  structure constant and strong interaction in Th-229}},
  \href{https://doi.org/10.1103/PhysRevA.73.034101}{\emph{Phys. Rev. A}
  {\bfseries 73} (2006) 034101}
  [\href{https://arxiv.org/abs/physics/0604188}{{\ttfamily physics/0604188}}].

\bibitem{Fadeev2020}
P.~Fadeev, J.C.~Berengut and V.V.~Flambaum, \emph{Sensitivity of
  $^{229}\mathrm{Th}$ nuclear clock transition to variation of the
  fine-structure constant},
  \href{https://doi.org/10.1103/PhysRevA.102.052833}{\emph{Phys. Rev. A}
  {\bfseries 102} (2020) 052833}.

\bibitem{2010JCAP...01..030N}
M.~{Nakashima}, K.~{Ichikawa}, R.~{Nagata} and J.~{Yokoyama},
  \emph{{Constraining the time variation of the coupling constants from cosmic
  microwave background: effect of {\ensuremath{\Lambda}}$_{QCD}$}},
  \href{https://doi.org/10.1088/1475-7516/2010/01/030}{\emph{\jcap} {\bfseries
  2010} (2010) 030} [\href{https://arxiv.org/abs/0910.0742}{{\ttfamily
  0910.0742}}].

\bibitem{Martins2021}
C.J.A.P.~{Martins}, \emph{{Primordial nucleosynthesis with varying fundamental
  constants. Degeneracies with cosmological parameters}},
  \href{https://doi.org/10.1051/0004-6361/202039605}{\emph{\aap} {\bfseries
  646} (2021) A47} [\href{https://arxiv.org/abs/2012.10505}{{\ttfamily
  2012.10505}}].

\bibitem{Damour2002}
T.~{Damour}, F.~{Piazza} and G.~{Veneziano}, \emph{{Violations of the
  equivalence principle in a dilaton-runaway scenario}},
  \href{https://doi.org/10.1103/PhysRevD.66.046007}{\emph{\prd} {\bfseries 66}
  (2002) 046007} [\href{https://arxiv.org/abs/hep-th/0205111}{{\ttfamily
  hep-th/0205111}}].

\bibitem{Vacher2023}
L.~{Vacher}, N.~{Sch{\"o}neberg}, J.D.F.~{Dias}, C.J.A.P.~{Martins} and
  F.~{Pimenta}, \emph{{Runaway dilaton models: Improved constraints from the
  full cosmological evolution}},
  \href{https://doi.org/10.1103/PhysRevD.107.104002}{\emph{\prd} {\bfseries
  107} (2023) 104002} [\href{https://arxiv.org/abs/2301.13500}{{\ttfamily
  2301.13500}}].

\bibitem{Lucca2024}
M.~{Lucca}, J.~{Chluba} and A.~{Rotti}, \emph{{CRRfast: an emulator for the
  cosmological recombination radiation with effects from inhomogeneous
  recombination}}, \href{https://doi.org/10.1093/mnras/stae915}{\emph{\mnras}
  {\bfseries 530} (2024) 668}
  [\href{https://arxiv.org/abs/2306.08085}{{\ttfamily 2306.08085}}].

\bibitem{Hart2023}
L.~{Hart} and J.~{Chluba}, \emph{{Using the cosmological recombination
  radiation to probe early dark energy and fundamental constant variations}},
  \href{https://doi.org/10.1093/mnras/stac3697}{\emph{\mnras} {\bfseries 519}
  (2023) 3664} [\href{https://arxiv.org/abs/2209.12290}{{\ttfamily
  2209.12290}}].

\bibitem{ACT}
S.~{Aiola}, E.~{Calabrese}, L.~{Maurin}, S.~{Naess}, B.L.~{Schmitt},
  M.H.~{Abitbol} et~al., \emph{{The Atacama Cosmology Telescope: DR4 maps and
  cosmological parameters}},
  \href{https://doi.org/10.1088/1475-7516/2020/12/047}{\emph{\jcap} {\bfseries
  2020} (2020) 047} [\href{https://arxiv.org/abs/2007.07288}{{\ttfamily
  2007.07288}}].

\bibitem{SPT}
J.T.~{Sayre}, C.L.~{Reichardt}, J.W.~{Henning}, P.A.R.~{Ade}, A.J.~{Anderson},
  J.E.~{Austermann} et~al., \emph{{Measurements of B -mode polarization of the
  cosmic microwave background from 500 square degrees of SPTpol data}},
  \href{https://doi.org/10.1103/PhysRevD.101.122003}{\emph{\prd} {\bfseries
  101} (2020) 122003} [\href{https://arxiv.org/abs/1910.05748}{{\ttfamily
  1910.05748}}].

\bibitem{SimonsObservatory}
{The Simons Observatory collaboration}, \emph{{The Simons Observatory}},  in
  \emph{BAAS}, vol.~51, p.~147, Sept., 2019
  [\href{https://arxiv.org/abs/1907.08284}{{\ttfamily 1907.08284}}].

\bibitem{LiteBIRD}
{LiteBIRD Collaboration}, E.~{Allys}, K.~{Arnold}, J.~{Aumont}, R.~{Aurlien},
  S.~{Azzoni} et~al., \emph{{Probing cosmic inflation with the LiteBIRD cosmic
  microwave background polarization survey}},
  \href{https://doi.org/10.1093/ptep/ptac150}{\emph{Progress of Theoretical and
  Experimental Physics} {\bfseries 2023} (2023) 042F01}
  [\href{https://arxiv.org/abs/2202.02773}{{\ttfamily 2202.02773}}].

\bibitem{DESI}
A.~Dey, D.J.~Schlegel, D.~Lang, R.~Blum, K.~Burleigh, X.~Fan et~al.,
  \emph{Overview of the desi legacy imaging surveys},
  \href{https://doi.org/10.3847/1538-3881/ab089d}{\emph{The Astronomical
  Journal} {\bfseries 157} (2019) 168}.

\bibitem{Euclid}
{Euclid Collaboration}, Y.~{Mellier}, {Abdurro'uf}, J.A.~{Acevedo Barroso},
  A.~{Ach{\'u}carro}, J.~{Adamek} et~al., \emph{{Euclid. I. Overview of the
  Euclid mission}},
  \href{https://doi.org/10.48550/arXiv.2405.13491}{\emph{arXiv e-prints} (2024)
  arXiv:2405.13491} [\href{https://arxiv.org/abs/2405.13491}{{\ttfamily
  2405.13491}}].

\bibitem{HIRES}
J.~Liske, G.~Bono, J.~Cepa et~al., \emph{{Top Level Requirements For
  ELT-HIRES}},  Tech. Rep. Document ESO 204697 Version 1 (2014).

\bibitem{burgess2007standard}
C.~Burgess and G.~Moore, \emph{The Standard Model: A Primer}, Cambridge books
  online, Cambridge University Press (2007).

\bibitem{Duff2001}
M.J.~{Duff}, L.B.~{Okun} and G.~{Veneziano}, \emph{{Trialogue on the number of
  fundamental constants}},
  \href{https://doi.org/10.1088/1126-6708/2002/03/023}{\emph{Journal of High
  Energy Physics} {\bfseries 2002} (2002) 023}
  [\href{https://arxiv.org/abs/physics/0110060}{{\ttfamily physics/0110060}}].

\bibitem{Duff2002}
M.J.~{Duff}, \emph{{Comment on time-variation of fundamental constants}},
  \href{https://doi.org/10.48550/arXiv.hep-th/0208093}{\emph{arXiv e-prints}
  (2002) hep} [\href{https://arxiv.org/abs/hep-th/0208093}{{\ttfamily
  hep-th/0208093}}].

\bibitem{Ellis2005}
G.F.R.~{Ellis} and J.-P.~{Uzan}, \emph{{c is the speed of light, isn't it?}},
  \href{https://doi.org/10.1119/1.1819929}{\emph{American Journal of Physics}
  {\bfseries 73} (2005) 240}
  [\href{https://arxiv.org/abs/gr-qc/0305099}{{\ttfamily gr-qc/0305099}}].

\bibitem{Duff2015}
M.J.~{Duff}, \emph{{How fundamental are fundamental constants?}},
  \href{https://doi.org/10.1080/00107514.2014.980093}{\emph{Contemporary
  Physics} {\bfseries 56} (2015) 35}
  [\href{https://arxiv.org/abs/1412.2040}{{\ttfamily 1412.2040}}].

\bibitem{Rich2015}
J.~{Rich}, \emph{{Which fundamental constants for cosmic microwave background
  and baryon-acoustic oscillation?}},
  \href{https://doi.org/10.1051/0004-6361/201526847}{\emph{\aap} {\bfseries
  584} (2015) A69} [\href{https://arxiv.org/abs/1503.06012}{{\ttfamily
  1503.06012}}].

\bibitem{Uzan2024}
J.-P.~Uzan, \emph{{Fundamental constants: from measurement to the universe, a
  window on gravitation and cosmology}},
  \href{https://arxiv.org/abs/2410.07281}{{\ttfamily 2410.07281}}.

\bibitem{Kobayashi2022}
T.~{Kobayashi}, A.~{Takamizawa}, D.~{Akamatsu}, A.~{Kawasaki}, A.~{Nishiyama},
  K.~{Hosaka} et~al., \emph{{Search for Ultralight Dark Matter from Long-Term
  Frequency Comparisons of Optical and Microwave Atomic Clocks}},
  \href{https://doi.org/10.1103/PhysRevLett.129.241301}{\emph{\prl} {\bfseries
  129} (2022) 241301} [\href{https://arxiv.org/abs/2212.05721}{{\ttfamily
  2212.05721}}].

\bibitem{Brans1961}
C.~Brans and R.H.~Dicke, \emph{Mach's principle and a relativistic theory of
  gravitation}, \href{https://doi.org/10.1103/PhysRev.124.925}{\emph{Phys.
  Rev.} {\bfseries 124} (1961) 925}.

\bibitem{Magueijo2003}
J.~{Magueijo}, \emph{{New varying speed of light theories}},
  \href{https://doi.org/10.1088/0034-4885/66/11/R04}{\emph{Reports on Progress
  in Physics} {\bfseries 66} (2003) 2025}
  [\href{https://arxiv.org/abs/astro-ph/0305457}{{\ttfamily
  astro-ph/0305457}}].

\bibitem{ATLAS:2023lhg}
{\scshape ATLAS} collaboration, \emph{{A precise determination of the
  strong-coupling constant from the recoil of $Z$ bosons with the ATLAS
  experiment at $\sqrt{s} = 8$ TeV}},
  \href{https://arxiv.org/abs/2309.12986}{{\ttfamily 2309.12986}}.

\bibitem{PDG2024}
{\scshape Particle Data Group} collaboration, \emph{{Review of Particle
  Physics} [to be published]}, {\emph{Phys. Rev. D} {\bfseries 110} (2024)
  030001}.

\bibitem{Touboul2022}
P.~{Touboul}, G.~{M{\'e}tris}, M.~{Rodrigues}, J.~{Berg{\'e}}, A.~{Robert},
  Q.~{Baghi} et~al., \emph{{M I C R O S C O P E Mission: Final Results of the
  Test of the Equivalence Principle}},
  \href{https://doi.org/10.1103/PhysRevLett.129.121102}{\emph{\prl} {\bfseries
  129} (2022) 121102} [\href{https://arxiv.org/abs/2209.15487}{{\ttfamily
  2209.15487}}].

\bibitem{Dent2007}
T.~{Dent}, \emph{{Composition-dependent long range forces from varying
  m$_{p}$/m$_{e}$}},
  \href{https://doi.org/10.1088/1475-7516/2007/01/013}{\emph{\jcap} {\bfseries
  2007} (2007) 013} [\href{https://arxiv.org/abs/hep-ph/0608067}{{\ttfamily
  hep-ph/0608067}}].

\bibitem{Berge2023}
J.~Bergé, \emph{Microscope’s view at gravitation},
  \href{https://doi.org/10.1088/1361-6633/acd203}{\emph{Reports on Progress in
  Physics} {\bfseries 86} (2023) 066901}.

\end{thebibliography}\endgroup
\bibliographystyle{JHEP}

\appendix
\section{Mass variations and their physicality \label{app:mass_phys}}

In this section we discuss the physicality of considering variations in individual masses (as opposed to the ratios of two different masses) in \cref{app:ssec:massratios} as well as arguing about variations of physical constants more generally in \cref{app:ssec:constants}. We begin with a foreword on mass variations more generally in \cref{app:ssec:masses}

\subsection{Foreword about mass variations}\label{app:ssec:masses}

As displayed in \cref{tab:fundamental}, our standard model of particle physics and general relativity contains 19 fundamental constants (excluding neutrino mass parameters and unit defining constants).\footnote{The review of \cite{Uzan2011} displays the same number of dimensionless constants by considering $\theta_{\rm QCD}$. Since whether this constant is a necessary addition to the standard model is not yet settled, we decided not to include it in our count.} The masses of the fundamental particles $m_x$ are not considered as part of this list, but can be derived from the combination of the Yukawa couplings $h_x$ and the Higgs vacuum expectation value $v$ as $m_x = h_x v/2$. On the other hand, the masses of composite particles, such as all hadrons, is mostly set by the energy of the interaction between its constituents and involve the value of the gauge couplings -- Effectively the masses of protons and neutrons, for example, is given by the interaction strength of the strong nuclear force with only minor corrections from the lightest quark masses. 

Therefore, the particle masses are always the byproduct of interactions between more fundamental fields. Nevertheless, our standard model predicts that all masses are constant over cosmological timescales \cite{burgess2007standard}. As such, they can be seen as direct proxies of the other fundamental constants contained in \cref{tab:fundamental} while having a more easily measurable and interpretable impact on the majority of the cosmological observables. A variation of these masses, or something mimicking this effect, would be a direct signature of new physics, implying the existence of yet unknown phenomena at play in the Higgs sector or within the gauge interactions. We discuss such models in \cref{sec:physical}. 

\subsection{Individual mass variations}\label{app:ssec:massratios}

It is widely acknowledged that only the space-time variations of dimensionless quantities can be investigated and constrained \cite{Duff2001,Duff2002,Ellis2005,Uzan2011,Duff2015,Rich2015, Uzan2024}. For identical reasons that one would want to look at gauge invariant quantities in field theories, dimensionless ratios are indeed independent of any choice of unit and are expected to provide absolute scales to quantify the intensity of physical phenomena.

Instead, particle masses are dimensional quantities. As such, past literature typically preferred dimensionless ratios between different particle masses, such as the proton to electron mass ratio 
\begin{equation}
\mu = \frac{m_p}{m_e}
\end{equation}
indeed because it is also commonly appearing in considerations of atomic energy levels, such as the amplitude of the fine-structure splitting. See also \cite{Kobayashi2022,Rich2015} for a discussion on this. Since we are only varying one mass while keeping others fixed, in our case varying the ratio is effectively equivalent to varying the individual mass that we do not keep fixed.

\subsection{Unit-defining constants}\label{app:ssec:constants}
Compared to mass variations, we do not argue about unit-defining constants, which are also listed in \cref{tab:fundamental} for completeness. Not only can arbitrary new units be defined using new constants (e.g. candela) but their variation is intrinsically not measurable \cite{Duff2002,Uzan2011}. This is also evident by the lack of an uncertainty to their definition in \cref{tab:fundamental}. Therefore, variations in these unit-defining constants per-se are just unit-redefinitions and as such unphysical.

However, it is important to stress that the associated concepts could be changed instead -- for the gravitational constant one can change the laws of gravitational coupling (e.g. scalar-tensor theories and especially Brans-Dicke theories \cite{Brans1961}) or for the speed of light the physics of light-propagation, see for example \cite{Ellis2005} for an excellent description of the different concepts that usually coincide as the \enquote{speed of light} (hence our \cref{tab:fundamental} lists the fundamental concept as a \enquote{spacetime constant} instead) and \cite{Magueijo2003} for a review of consistent varying speed of light models.

\begin{table*}[t]
    \begin{threeparttable}
    \centering
    \resizebox{\columnwidth}{!}{\begin{tabular}{c c | c c }
        Constant & Symbol & Value & Uncertainty \\
        \hline \multicolumn{4}{c}{True constants}\\ \hline
        Strong coupling constant\tnote{*}  -- $SU(3)$&$\alpha_s$ &  0.1183 & 0.0009 \\
        Electromagnetic coupling  constant -- $SU(2)$+ $U(1)$ &$\alpha_\mathrm{EM}^{-1}$ & 137.035999084 & 0.000000021 \\
        Weinberg angle\tnote{*} -- $SU(2)$+ $U(1)$ &$\sin^2(\theta_W)$ &  0.23120 & 0.00015 \\ \hline
        Higgs VEV & $v$ & 246.21964GeV & 0.00006GeV  \\
        Higgs mass & $m_\mathrm{Higgs}$ & 125.22GeV & 0.14GeV\\ \hline
        Electron Yukawa & $h_e$ & $2.94 \times 10^{-6}$ & $7 \cdot 10^{-13}$ \\ 
        Muon Yukawa & $h_\mu$ & $6.07\cdot 10^{-4}$ & $1.3\cdot 10^{-11}$ \\ 
        Tau Yukawa & $h_\tau$ & 0.01020617 & 0.00000052 \\ 
        Down Yukawa & $h_d$ &$2.70\cdot 10^{-5}$ & $4.0 \cdot 10^{-7}$\\ 
        Up Yukawa & $h_u$ & $1.24\cdot 10^{-5}$ & $4.0 \cdot 10^{-7}$\\ 
        Strange Yukawa & $h_s$ & $5.37\cdot 10^{-4}$ & $4.6\cdot 10^{-6}$\\ 
        Charm Yukawa & $h_c$ & $7.31\cdot 10^{-3}$ & $2.64\cdot 10^{-6}$ \\ 
        Bottom Yukawa & $h_b$ & 0.024 & $4.0\cdot 10^{-5}$ \\ 
        Top Yukawa & $h_t$ & $0.991$ & $0.002$ \\ \hline 
        Quark CKM & $\sin \theta_{12}$ & 0.22501  & 0.00068 \\
        Quark CKM & $\sin \theta_{23}$ & 0.04183 & ${}^{+0.00079}_{-0.00069}$\\
        Quark CKM & $\sin \theta_{13}$ & 0.003732 & 0.000087\\
        Quark CKM phase & $\delta_\mathrm{CKM}$ & 1.147 & 0.026\\ \hline
        Neutrino PMNS\tnote{§} & $ \sin^2 \theta_{12}$ & 0.30 & 0.01\\
        Neutrino PMNS\tnote{§} & $ \sin^2 \theta_{23}$ & 0.57 & 0.02\\
        Neutrino PMNS\tnote{§} & $ \sin^2 \theta_{13}$ & 0.220 & 0.006\\
        Neutrino PMNS phase\tnote{§} & $\delta_\mathrm{CP}$ & 197${}^{\circ}$ & ${}^{+42}_{-25}$${}^{\circ}$\\
        Neutrino mass\tnote{§} & $\Delta m^2_\mathrm{sol}$ & $7.41 \cdot 10^{-5} eV^2$& $0.20 \cdot 10^{-5} eV^2$\\
        Neutrino mass\tnote{§} & $\Delta m^2_\mathrm{atm}$ &  $2.437 \cdot 10^{-3} eV^2$& $0.028 \cdot 10^{-3} eV^2$\\
        Lightest Neutrino mass\tnote{§} & $m_\mathrm{lightest}$ & Not yet measured & Not yet measured \\
        \hline
        Gravitational constant & $G$ & 6.6743 $\cdot 10^{-11}$ $\mathrm{kg}^{-1}\mathrm{m}^3 \mathrm{s}^{-2}$ & 1.5 $\cdot 10^{-15}$ $\mathrm{kg}^{-1}\mathrm{m}^3 \mathrm{s}^{-2}$\\
        \hline \multicolumn{4}{c}{Unit definitions}\\ \hline
        Cs hyperfine frequency & $\Delta \nu_\mathrm{Cs}$ & 9192631770 Hz & ---\\
        Elementary charge & $e$ & 1.602176634 $\cdot 10^{-19}$ C & ---\\
        Boltzmann constant & $k_B$ & 1.380649 $\cdot 10^{-23}$ J/K & ---\\
        Spacetime constant & $c$ & 299792458 m/s & ---\\
        Planck constant & $h$ & 6.62607015 $10^{-34}$ J/Hz & ---\\
        \ldots & \ldots & \ldots  & ---
    \end{tabular}}
\caption{All values except for the strong coupling constant (which is from \cite{ATLAS:2023lhg}) are from \cite{PDG2024}. One can also rephrase the weak ($SU(2)+U(1)$) constants through the equivalent set of $g\sim 0.65$ for $SU(2)$ and $g' \sim 0.35$ for $U(1)$, which are related by $\sin(\theta_W) = g'/\sqrt{g'^2+g^2}$ and $\alpha_\mathrm{EM}^{-1}(E=M_Z) = g^2 \cdot \sin(\theta_W)$, where $\alpha_\mathrm{EM}^{-1}$ and $\alpha_\mathrm{EM}^{-1}(E=M_Z) \simeq 127.918 \pm 0.018$ are related through the usual renormalization group equations. We also have the Fermi constant $G_F = 1/(\sqrt{2} v^2)$ and the Higgs constants are $m_\mathrm{Higgs} = \sqrt{-\mu_\mathrm{Higgs}^2/2}$ and $v = m_\mathrm{Higgs} \sqrt{2/\lambda_\mathrm{Higgs}}$ with $\mu_\mathrm{Higgs}^2$ being the quadratic and $\lambda_\mathrm{Higgs}$ the quartic Higgs coupling. It is worth mentioning that the only dimensionful constants are the Higgs VEV ($v$, the mass is related through a dimensionless factor) and the gravitational constant, which means that the numerical value for these both are tightly related to the hyperfine frequency $\Delta \nu_\mathrm{Cs}$ used to define the second. All Fermion masses are simply  $m_x = h_x v /\sqrt{2}$. For the neutrino properties, it is possible that the number of parameters can change depending on a future improved understanding of the origin of their oscillations.}
    \label{tab:fundamental}
    \begin{tablenotes}
        \item [*] Measured at the Z-boson mass-equivalent energy of $M_Z=91.2$GeV.
        \item [§] These quantities relating to the neutrino sector are not usually included in the\\ standard model of particle physics. We present the fit values for normal ordering,\\ see \cite{PDG2024} for other values.
    \end{tablenotes}
    \end{threeparttable}
    \vspace{1\baselineskip}
\end{table*}

\section{Masses and the universality of free fall \label{app:EEP}}

The deviation to the universality of free fall is quantified by the Eötvos parameter comparing the acceleration of two free falling bodies $A$ and $B$:

\begin{equation}
    \eta(A,B) = 2\frac{a_A-a_B}{a_A+a_B}~.
\end{equation}

Within general relativity, the Einstein Equivalence Principle (EEP) is true as an elementary axiom, hence $a_A=a_B$ and $\eta(A,B)=0\,\forall A,B$. However, this relationship can be broken in extended theories of gravity or theories of varying fundamental constants. This is because variations of fundamental constants can induce time-dependent changes of the inertial mass. 

Greatly simplifying, since ${\rm d}\vec{p}/{\rm d}t = m(t)\vec{g}$ with $\vec{p}=m(t)\vec{v}$, one obtains
\begin{equation}
    \frac{{\rm d}\ln(m)}{{\rm d} t} v + a = g~,
\end{equation}
and we observe that there is an additional effective force from mass variations. 
The mass of test atoms $A$ and $B$ is given in general by
\begin{equation}
m_i = (A_i-Z_i) m_n + Z_i m_p + Z_i m_e - E_B^\mathrm{nuclear}-E_B^\mathrm{atomic}~,
\end{equation}
where $i \in \{A,B\}$, $A_i$ is the atomic mass number, $Z_i$ is the proton number, and $E_B$ is the binding energy (nuclear or atomic).
Therefore, when considering exclusively a time variation of $m_e$ (no other coupled constant variations)
\begin{equation}
\frac{{\rm d}\ln\left(m_i\right)}{{\rm d}t}  = \frac{Z_i - \mathrm{d}(E_{i,B}^\mathrm{atomic})/\mathrm{d}m_e}{m_i} \,\frac{{\rm d} m_e}{{\rm d}t }~,
\end{equation} 
which shows that in principle such electron mass variations would induce a non-universal free-fall for non-ionized atoms. 

In principle, it is not unfeasible to compute the dependence of the atomic binding energies of common elements used in free-fall experiments (such as MICROSCOPE \cite{Touboul2022}, which uses Ti and Pt) and achieves $\eta({\rm Ti},{\rm Pt})=\left(-1.5 \pm 2.3 (\,{\rm stat}) \pm 1.5 (\,{\rm syst})\right) \times 10^{-15}$. 

One can estimate roughly what order of magnitude such constraints should translate to. Approximating $\mathrm{d}\ln m_i/\mathrm{d}t \approx \mathrm{const}$ during the timescales relevant for the experiment, we find that the $\eta \propto \Delta \mathrm{d}\ln m_i/\mathrm{d}t \cdot \Delta t$, where $\Delta t$ is the duration of the experiment. In that case, we would approximate that for a $\mathcal{O}(\mathrm{year})$-long duration of MICROSCOPE, one would have constraints on $\Delta \mathrm{d}\ln m_i/\mathrm{d} \ln a \sim 10^{-5}$. Now neglecting for the rough approximation the $\mathrm{d}E_{i,B}^\mathrm{atomic}/\mathrm{d}m_e$ term, we get $\Delta \mathrm{d}\ln m_i/\mathrm{d} \ln a \sim \Delta (Z_i/m_i) \cdot m_e \cdot  \mathrm{d}\ln m_e/\mathrm{d} \ln a$ (and $Z_\mathrm{Ti}=22$, $m_\mathrm{Ti}=47.9\,\mathrm{amu}$, $Z_\mathrm{Pt}=78$, $m_\mathrm{Pt}=195.1\,\mathrm{amu}$, which gives $\Delta (Z_i/m_i) \cdot m_e \simeq 3\cdot 10^{-5}$), we can translate this into a bound on $\mathrm{d}\ln m_e/\mathrm{d}\ln a$ of the order $\mathcal{O}(1)$, which would not be competitive with atomic clock bounds, which are at the level $10^{-7}$.
More typically, coupled variations of the constants will be present in a given model and thus the bound will have to be derived within the context of that given specific model (leading to much more competitive bounds). A more detailed discussion on how universality of free fall bounds can be applied to specific models can be found in \cite{Dent2007,Damour1994,Berge2023}. Due to this model-dependence we do not use these bounds in the main work.

\end{document}